\documentclass[twocolumn,prb,aps,floats,floatfix,showpacs,amsmath,amssymb,superscriptaddress]{revtex4-1}
\usepackage{graphicx}
\usepackage{bm}
\usepackage{times}
\usepackage{amssymb} 
\usepackage{amsmath}
\usepackage{mathrsfs} 

\def\Xint#1{\mathchoice
{\XXint\displaystyle\textstyle{#1}}%
{\XXint\textstyle\scriptstyle{#1}}%
{\XXint\scriptstyle\scriptscriptstyle{#1}}%
{\XXint\scriptscriptstyle\scriptscriptstyle{#1}}%
\!\int}
\def\XXint#1#2#3{{\setbox0=\hbox{$#1{#2#3}{\int}$ }
\vcenter{\hbox{$#2#3$ }}\kern-.5\wd0}}

\def\dashint{\Xint-}

\begin{document}

\title{Decoherence by electromagnetic fluctuations in
double-quantum-dot charge qubits}

\author{Diego C. B. Valente}

\affiliation{Department of Physics, University of Connecticut, Storrs,
  Connecticut 06269, USA,}

\author{Eduardo R. Mucciolo}

\affiliation{Department of Physics, University of Central Florida,
  Orlando, Florida 32816-2385, USA,}

\author{F. K. Wilhelm}

\affiliation{Institute of Quantum Computation and Department of
  Physics and Astronomy, University of Waterloo, Waterloo, Ontario,
  Canada N2L 3G1}

\date{\today}

\begin{abstract}
We discuss decoherence due to electromagnetic fluctuations in charge
qubits formed by two lateral quantum dots. We use effective circuit
model to evaluate correlations of voltage fluctuations in the qubit
setup. These correlations allows us to estimate energy ($T_1$) and
phase ($T_2$) relaxation times of the qubit system. Our theoretical
estimate of the quality factor due to dephasing by electromagnetic
fluctuations yields values much higher than those found in recent
experiments, indicating that other sources of decoherence play a
dominant role.
\end{abstract}

\pacs{73.21.La, 03.67.Lx, 73.23.Hk}

\maketitle

\section{Introduction}

Solid-state semiconductor lateral quantum dots are strong candidates
for the physical realization of qubits. These artificial systems can
be designed to allow for the observation of coherent oscillations
between their quantum states. Since its first
proposals,\cite{loss98,blick00} a wide variety of experiments have
demonstrated control over the spin degree of freedom of confined
electrons in quantum dots,\cite{petta05,koppens06} as well as charge
states.\cite{tanamoto00,hayashi03,petta04,gorman05} Solid-state
quantum computer architectures with qubits encoded in dopant atoms in
semiconductor crystals have also been proposed.\cite{hollenberg04}
Quantum dots present the ubiquitous advantages of being manufactured
from highly developed semiconductor technology and may offer easier
scalability, the latter being key in enabling the manufacturing of
large-scale quantum computers in the future. A drawback to their use
in quantum computers is that they also couple rather effectively to
external degrees of freedom which lead to decoherence.

Semiconductor qubits are susceptible to various decoherence
mechanisms. Hyperfine coupling to lattice nuclear spins reduces the
phase coherence of electron spins,\cite{johnson,taylor} while qubits
based on the charge degree of freedom are particularly sensitive to
decoherence mechanisms related to charge motion, such as coupling to
phonon modes and to charge traps in the substrate. Several of these
sources have been investigated.\cite{fedichkin04, kempe, brandes02,
  sergueietaldoubledot05,wu04,Hu05,oi05,nonmarkovian,storcz05,
  hohenester07}

\begin{figure}[ht]
\centering
\includegraphics[width=8cm]{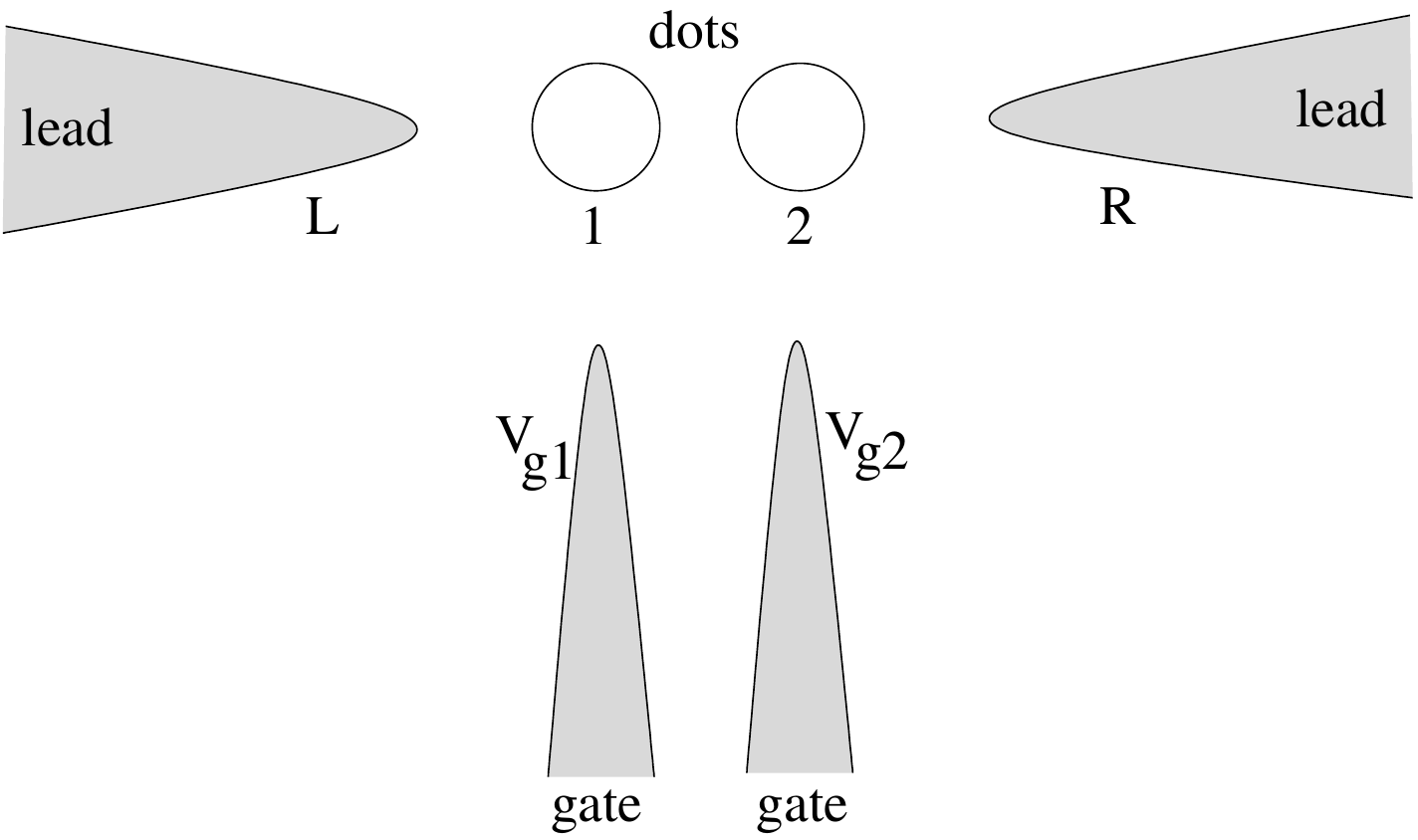}
\caption{(Color online) Schematic representation of a double quantum
dot setup.}
\label{fig:DQD}
\end{figure}

In this paper we focus on charge-based qubits. The simplest
realization of a charge qubit is a doubl-quantum-dot (DQD) system
with an odd number of electrons, as shown schematically in
Fig.~\ref{fig:DQD}. So far measurements of quality ($Q$) factors of
coherent oscillations in these systems have yielded rather low values
in the range of 3--9.\cite{hayashi03,petta04,gorman05} In an effort to
identify the main sources of decoherence, theoretical estimates of the
$Q$ factor have been carried out assuming mainly the coupling to
acoustic phonons.\cite{fedichkin04, kempe, brandes02,
  sergueietaldoubledot05,wu04,Hu05,oi05,nonmarkovian,storcz05,
  hohenester07} However, a discrepancy of at least one order of
magnitude remains between the experimental value and the theoretical
estimates, with the latter predicting larger $Q$ factors. This
discrepancy indicates that the phonons may not be the dominant noise
source in current experimental setups. Thus, an investigation of other
possible environmental decoherence mechanisms is in order. Here, we
consider the coupling of the DQD charge-based qubit systems to voltage
fluctuations in the gates.

This paper is organized as follows. In Sec.~\ref{sec:Doubledot} we
define the DQD effective Hamiltonian and the interaction between the
qubit and the gate-voltage fluctuations. In Sec.~\ref{sec:transmline}
we introduce the effective circuit model that describes the DQD and
the electromagnetic environment, as well as the Hamiltonian of the
latter. In Sec.~\ref{sec:Estimates} we estimate the equivalent circuit
parameters and in Sec.~\ref{sec:rates} we calculate upper bounds to
the decoherence rates and $Q$ factors. Our main finding is that
voltage fluctuations cause only a very small decoherence effect in DQD
charge qubits. Since double-dot {\it spin} qubits are also susceptible
to decoherence due to charge motion, in Sec. \ref{sec:spinqubits} we
use results from our circuit model of electromagnetic fluctuations to
estimate decoherence rates for those system. Our conclusions are
presented in Sec.~\ref{sec:conclusions}.

\section{Hamiltonian of the double-quantum-dot system}
\label{sec:Doubledot}

The Hamiltonian of a DQD can be separated into a quantum part related
to the occupation of energy levels on each dot and a classical part
that quantifies the charging energy,
\begin{eqnarray}
H = \sum_n \varepsilon_{1n}\, c_{1n}^\dagger c_{1n} + \sum_n
\varepsilon_{2n}\, c_{2n}^\dagger c_{2n} + E(N_1,N_2),
\end{eqnarray}
where $c^\dagger_{in}$ and $c_{in}$ are creation and annihilation
operators of the state with energy $\varepsilon_{in}$ in the left
($i=1$) or right dot ($i=2$). The dot occupation numbers are defined
as $N_i = \sum_n c_{in}^\dagger c_{in}$ while the total charging
energy is given by \cite{dutch}
\begin{eqnarray}
E(N_1,N_2) &=& \frac{E_{C1}}{2}N^2_1 + \frac{E_{C2}}{2}N^2_2 +
N_1N_2E_{Cm} \nonumber\\ && - \frac{1}{|e|}[C_{g1}V_{g1}(N_1E_{C1} +
  N_2E_{Cm})] \nonumber \\ && - \frac{1}{|e|}[C_{g2}V_{g2}(N_1E_{Cm} +
  N_2E_{C2})] \nonumber \\ && + \frac{1}{2e^2}[C^2_{g1}V^2_{g1}E_{C1}
  + C^2_{g2}V^2_{g2}E_{C2}] \nonumber\\ && +
\frac{1}{e^2}[C_{g1}V_{g1}C_{g2}V_{g2}E_{Cm}],
\label{energy}   
\end{eqnarray}
with the individual charging energies defined as
\begin{eqnarray}
E_{C1} &=& \frac{e^2}{C_1}\left( 1 - \frac{C^2_m}{C_1C_2}
\right)^{-1}, \\ E_{C2} &=& \frac{e^2}{C_2}\left( 1 -
\frac{C^2_m}{C_1C_2} \right)^{-1}, \\ E_{Cm} &=& \frac{e^2}{C_m}\left(
\frac{C_1C_2}{C^2_m} - 1 \right)^{-1}.
\label{chargenergy}
\end{eqnarray}
The capacitances and voltages shown in
Eqs. \eqref{energy}-\eqref{chargenergy} are defined in
Fig.~\ref{fig:DQDcirc}. $C_{1,2}$ is the sum of all capacitances
attached to dot 1 or 2: $C_{1,2} = C_{T1,2} + C_{g1,2} + C_m$.

For the purpose of our analysis, the Hamiltonian can be greatly
simplified. Notice that the DQD qubit can be viewed as a double-well
potential where an unpaired electron oscillates between both quantum
dots by tunneling through the potential barrier. Spin degrees of
freedom can be neglected. By adjusting the gate voltages, one can set
the system near the degeneracy point $E(1,0) = E(0,1)$, in which case
the logical states of the qubit correspond to the electron being on
the left or right, $|L \rangle$ ($N_1 = 1$ and $N_2 = 0$) and $|R
\rangle$ ($N_1 = 0$ and $N_2 = 1$), respectively. The typical
single-particle level spacing within each quantum dot is assumed
sufficiently large so that only one level on each dot needs to be
considered at low enough temperatures. The barrier height $\Delta$
determines the tunneling rate between the dots and can be adjusted by
a gate voltage while a bias $\varepsilon$ between the two dots can
also be applied through two independent plunger gate voltages. The
dynamics in the DQD qubit is then governed by the reduced two-level
Hamiltonian
\begin{eqnarray}
\label{eq:Hs}
H_S = \frac{\varepsilon}{2}\, \left( |L\rangle\langle L| -
|R\rangle\langle R| \right) + \frac{\Delta}{2} \left( |L\rangle\langle
R| + |R\rangle\langle L| \right),
\end{eqnarray}
with the constraint that $|L\rangle\langle L| + |R\rangle\langle R|=
1$. The fields $\epsilon$ and $\Delta$ represent the interdot bias and
the interdot capacitive coupling, respectively.

Electromagnetic noise is introduced into the DQD qubit system by means
of gate voltage fluctuations. These fluctuations may originate from
the voltage sources and the thermal noise in the transmission lines,
and introduce decoherence into the qubit system through interactions
with the electrons in the quantum dots. While the former can be
substantially reduced by careful filtering, the latter is less
controlled. Here we will focus on the noise coming from the plunger
gates. The effect of voltage fluctuations in the gate electrodes is
captured by the qubit-environment interaction
\begin{eqnarray}
\label{eq:Hnoise}
H_{SB} = e\eta \left( \delta V_{g1} - \delta V_{g2} \right)\, \left(
|L\rangle\langle L| - |R\rangle\langle R| \right),
\end{eqnarray}
where $\eta$ is the capacitive lever arm coefficient,
\begin{equation}
\eta = \frac{C_{g1}C_2 + C_{g2}C_1 - C_m ( C_{g1} + C_{g2})} {4(C_1C_2
  - C_m^2)}.
\end{equation}
Depending on the particular qubit setup, other sources of
electromagnetic noise may also exist, such as bias and current-voltage
fluctuations. They can affect not only the qubit coherent dynamics but
also the state measurement. For the sake of maintaining some
generality in our study, we will only treat electromagnetic
fluctuations which can be expressed as Eq. (\ref{eq:Hnoise}). In
addition, we will model the voltage fluctuations through
frequency-dependent impedances along the gate transmission lines.

\section{Hamiltonian for the electromagnetic environment}
\label{sec:transmline}

\begin{figure}[ht]
\centering
\includegraphics[width=7cm]{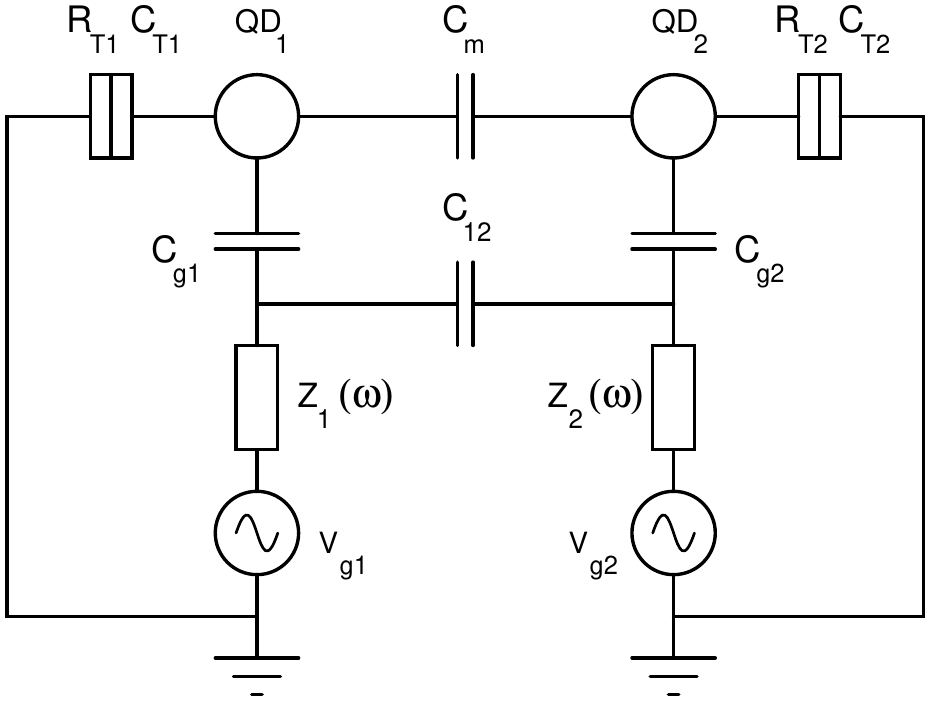}
\caption{Circuit representation of a double quantum dot system coupled
  to an electromagnetic environment through metallic gate
  electrodes. Source and drain electrodes are assumed grounded.  The
  quantum dots are denoted by QD$_1$ and QD$_2$.}
\label{fig:DQDcirc}
\end{figure}

The effective circuit of a double quantum dot setup is shown in
Fig.~\ref{fig:DQDcirc}. The effect of the electromagnetic environment
is modeled by the frequency-dependent impedances $Z_{1,2}(\omega)$. In
the experimental setups, the voltage lines typically run parallel to
each other over several microns or more. In order to take into account
any capacitive coupling between the lines, we introduced capacitance
$C_{12}$ into the circuit.

\begin{figure}[b]
\centering
\includegraphics[width=8cm]{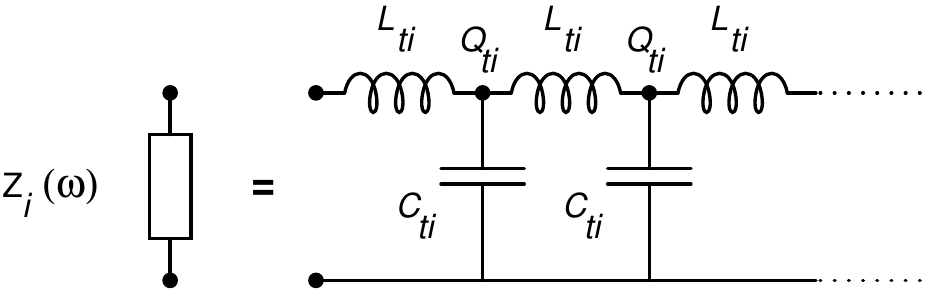}
\caption{Circuit representation of the electromagnetic environment as
  a transmission line.}
\label{fig:TL}
\end{figure}

The impedances $Z_{1,2}(\omega)$ can be modeled by means of a
transmission line with distributed elements, which stems from the fact
that the source of noise in our circuit is spatially distributed along
a finite length. Let us consider first each transmission line
independently, as shown in Fig.~\ref{fig:TL}, whose impedance
$Z_i(\omega)$ can be represented by an infinite ladder network of
identical inductors $L_{ti}$ and capacitors $C_{ti}$ (see
Ref. \onlinecite{feynman}),
\begin{eqnarray}
\label{TL_imped}
Z_i(\omega) = \frac{1}{2} \left( i \omega L_{ti} + \sqrt{ - \omega^2
  L_{ti}^2 + 4\frac{L_{ti}}{C_{ti}}} \right).
\end{eqnarray}
Typically, it would be necessary to estimate the values of the
spatially distributed resistance, capacitance, and inductance in the
circuit, but the choice to model the impedance as a $LC$ transmission
line can be made because it is known that through a (not necessarily
trivial) normal-mode transformation, any $RLC$ or $RC$ transmission
line can be written as an infinite LC ladder network. The elements
$C_{ti}$ and $L_{ti}$ of the transmission line can be determined from
two real parameters of the real: the cutoff frequency $\omega_{c}$ and
the low frequency asymptotic limit to the characteristic impedance
$Z(\omega = 0)$. In an semi-infinite line, ${\rm Re}\{Z(\omega)\} = 0$
when $\omega \ge \omega_c$. Hence,
\begin{eqnarray}
\label{TLwc}
\omega_{c} = \frac{2}{\sqrt{L_{ti}C_{ti}}}.
\end{eqnarray}
$Z_i(\omega = 0)$, on the other hand, can be calculated by taking the
low frequency asymptotic limit of Eq. \eqref{TL_imped}.  It is
straightforward to see that this limit yields
\begin{eqnarray}
Z_i(\omega = 0) = \sqrt{\frac{L_{ti}}{C_{ti}}} = R,
\label{TLparam}
\end{eqnarray}
where $R$ is an ohmic resistance.

To introduce noise, the transmission line is decomposed into normal
modes. In quantized form, the charge at the $l$th node, $Q_{l,i}$, and
the flux $\phi_{l,i}$ are conjugated variables obeying the commutation
relation $[\phi_{l,i},Q_{l^\prime,i^\prime}] = ie
\delta_{i,i^\prime}\delta_{l,l^\prime}$. Following the standard
procedure, we define the Hamiltonian governing the flux and charge
fluctuations along such transmission as
\begin{eqnarray}
H_{T,i} = \frac{Q^2_{0,i}}{2C_{gi}} +
\sum^{+\infty}_{l=1}\left[\frac{Q^2_{l,i}}{2C_{ti}} +
  \frac{\hbar^2}{e^2} \frac{(\phi_{l,i}-\phi_{l-1,i})^2}{2L_{ti}}
  \right].
\label{hamiltTL}  
\end{eqnarray}
Notice that $C_{gi}$ represents the capacitive coupling between the
quantum dots and their respective gates, while $C_{ti}$ and $L_{ti}$
represent the capacitive and inductive terms, respectively, at each
rung in the transmission line.

Adding the capacitive coupling between the voltage transmission lines,
we obtain the following environmental noise Hamiltonian:
\begin{eqnarray}
H_B = H_{T,1} + H_{T,2} + \frac{Q_{0,1} Q_{0,2}}{C_{12}}.
\end{eqnarray}
The cross term complicates the task of finding the normal models of
the environment and an alternative approach was adopted.

\section{Double-dot junction}

\begin{figure}[h]
\centering
\includegraphics[width=8cm]{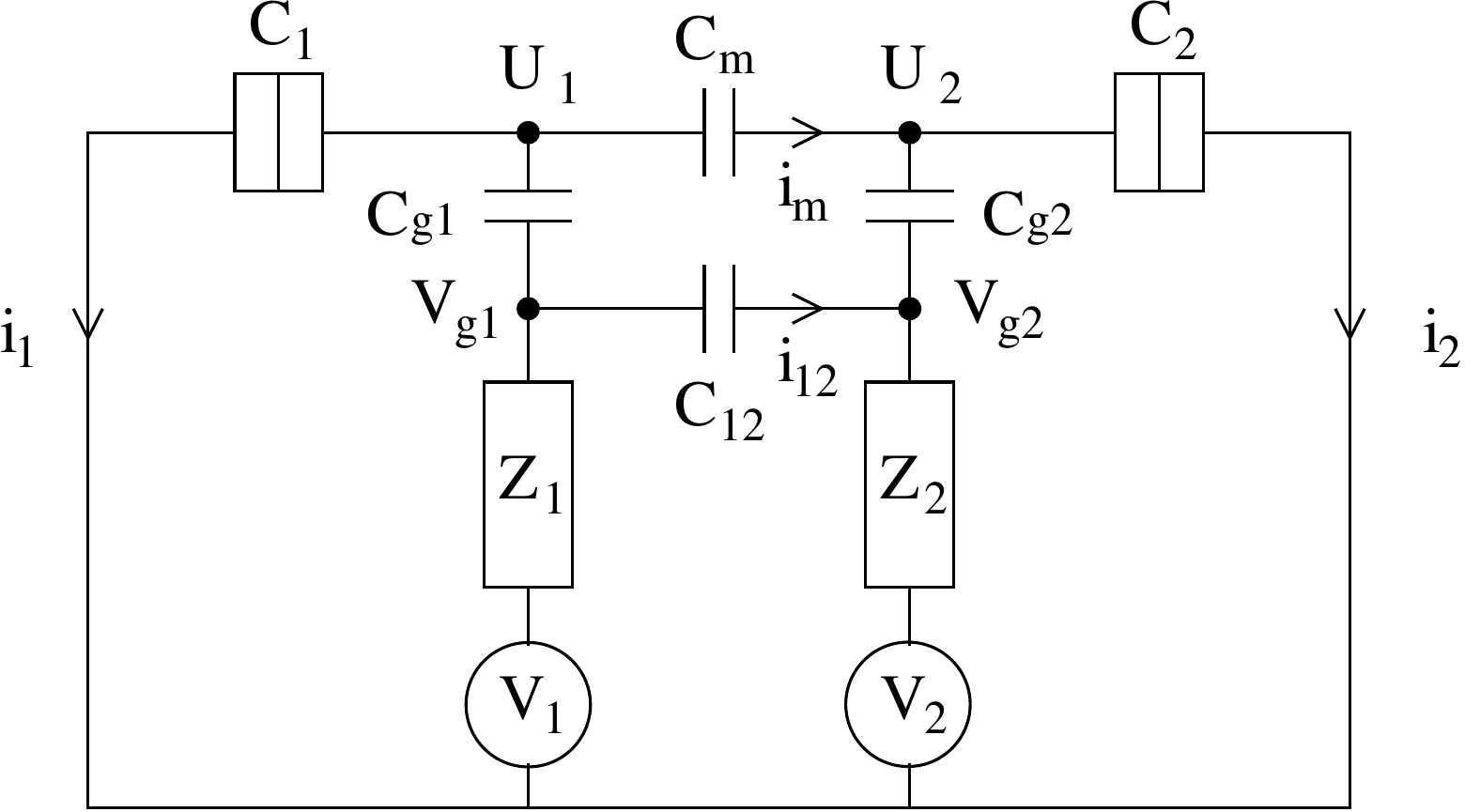}
\caption{Circuit of a double-dot junction system coupled to two
  voltage sources through noisy lines.}
\label{fig:2dot}
\end{figure}

The double junction solution is based on the original solution for a
single-dot junction treated in detail by Ingold and Nazarov.
\cite{ingoldnazarov} The nontrivial aspect of our extension of the
calculations in Ref. \onlinecite{ingoldnazarov} is the inclusion of
the gate capacitances (see Fig. \ref{fig:2dot}).

We start with the setup shown in Fig. \ref{fig:2dot}. Following a
straightforward application of Kirchhoff's laws, we find the relations
\begin{eqnarray}
\label{eq:V1}
V_1 & = & \left( i_1 + i_{12} + i_m \right)\, Z_1 + V_{g1}, \\
\label{eq:V2}
V_2 & = & \left( i_1 - i_{12} - i_m \right)\, Z_2 + V_{g2},
\end{eqnarray}
with
\begin{equation}
\label{eq:i12}
V_{g1} - V_{g2} = i_{12}\, Z_{12}
\end{equation}
and
\begin{equation}
\label{eq:im}
U_1 - U_2 = i_m\, Z_m,
\end{equation}
where $Z_{12} = \left( i\omega C_{12} \right)^{-1}$ and $Z_m = \left(
i\omega C_m \right)^{-1}$.

We begin by eliminating $i_{12}$ and $i_m$ in Eqs. (\ref{eq:V1}) and
(\ref{eq:V2}) with the help of Eqs. (\ref{eq:i12}) and (\ref{eq:im}),
and proceed to write $V_{g1}$ and $V_{g2}$ in terms of $U_1$ and
$U_2$. For this purpose, we notice that
\begin{eqnarray*}
\label{eq:Vgint_1}
V_{g1} & = & U_1 + \left( i_1 + i_m \right)\, Z_{g1}, \\
V_{g2} & = & U_2 + \left( i_2 - i_m \right)\, Z_{g2},
\end{eqnarray*}
$C_i\, U_i = Q_i$, and $i_1 = \dot{Q}_1$, with $i=1,2$. Eliminating
$V_{g1}, V_{g2}$ and rewriting $V_1, V_2$ in terms of $i_1, i_2, Q_1$
and $Q_2$, we obtain, in matrix notation
\begin{equation}
\label{eq:Vvector}
\left( \begin{array}{c} V_1 \\ V_2 \end{array} \right) = {\cal Z}
\cdot \left( \begin{array}{c} \dot{Q}_1 \\ \dot{Q}_2 \end{array}
\right) + \tilde{\cal C}^{-1} \cdot \left( \begin{array}{c} Q_1
  \\ Q_2 \end{array} \right),
\end{equation}
where
\begin{equation}
\label{eq:Z_cal}
{\cal Z} = \left( \begin{array}{cc} {\cal Z}_{11} & {\cal Z}_{12}
  \\ {\cal Z}_{21} & {\cal Z}_{22} \end{array} \right),
\end{equation}
with
\begin{widetext}
\begin{eqnarray}
\label{eq:Z_cal_comp}
{\cal Z}_{11} & = & Z_1 \left( 1 + \frac{C_m}{C_1} \right) + Z_1
C_{12} \left[ \frac{1}{C_1}\left( 1 + \frac{C_m}{C_{g1}} \right) +
  \frac{1}{C_{g1}} + \frac{C_m}{C_{g2}C_1} \right], \\ {\cal Z}_{12} &
    = & -Z_1\frac{C_m}{C_2} - Z_1 C_{12} \left[ \frac{C_m}{C_{g1}C_2} +
      \left( 1 + \frac{C_m}{C_{g2}} \right) \frac{1}{C_2} +
      \frac{1}{C_{g2}} \right], \\ {\cal Z}_{21} & = &
    -Z_2\frac{C_m}{C_1} - Z_2 C_{12} \left[ \frac{C_m}{C_{g2}C_1} +
      \left( 1 + \frac{C_m}{C_{g1}} \right) \frac{1}{C_1} +
      \frac{1}{C_{g1}} \right], \\ {\cal Z}_{22} & = & Z_2 \left( 1 +
      \frac{C_m}{C_2} \right) + Z_2 C_{12} \left[ \frac{1}{C_2}\left(
      1 + \frac{C_m}{C_{g2}} \right) + \frac{1}{C_{g2}} +
      \frac{C_m}{C_{g1}C_2} \right].
\end{eqnarray}
The matrix $\tilde{C}$ is defined as
\begin{equation}
\label{eq:C_tilde}
\tilde{\cal C} = \frac{1}{\det{(\tilde{\cal C}^{-1})}}
\left( \begin{array}{cc} \left[ 1 + \frac{C_m}{C_{g2}}
    \right]\frac{1}{C_2} + \frac{1}{C_{g2}} & \frac{C_m}{C_{g1} C_2}
  \\ \frac{C_m}{C_{g2} C_1} & \left[ 1 + \frac{C_m}{C_{g1}}
    \right]\frac{1}{C_1} + \frac{1}{C_{g1}} \end{array} \right),
\end{equation}
where 
\begin{equation}
\label{eq:C_tilde_inv_det}
\det{(\tilde{\cal C}^{-1})} = - \frac{C_m^2}{C_{g1}C_{g2}C_1C_2} +
\left[\left( 1 + \frac{C_m}{C_{g1}} \right)\frac{1}{C_1} +
\frac{1}{C_{g1}} \right] \left[ \left( 1 + \frac{C_m}{C_{g2}}
  \right)\frac{1}{C_2} + \frac{1}{C_{g2}} \right].
\end{equation}
\end{widetext}
Notice that when we set $C_m = 0$ and $C_{12} = 0$ in
Eq. (\ref{eq:Vvector}), we decouple the two halves of the circuit and
obtain two independent equations for each half of the circuit.

In analogy to the single dot-junction circuit, the Hamiltonian for the
environment in this case can be written as
\begin{eqnarray}
H_{\rm env} &=& H_{\rm charge} + \sum_{n=1}^\infty
\frac{q_{n1}^2}{2C_{n1}} + \left ( \frac{\hbar}{e} \right)^2
\frac{\left( \tilde{\varphi}_{g1} - \varphi_{n1} \right)^2}{2L_{n1}}
\nonumber\\ & & +\ \frac{ q_{n2}^2}{2C_{n2}} + \left ( \frac{\hbar}{e}
\right)^2 \frac{\left( \tilde{\varphi}_{g2} - \varphi_{n2} \right)^2
}{2L_{n2}}.
\end{eqnarray}
We find that
\begin{equation}
\dot{q}_{ni}(t) = G_{ni}(t) + \frac{\hbar}{e L_{ni}} \int_0^t
dt^\prime\, \cos \left[ \omega_{ni} (t-t^\prime) \right]\,
\dot{\tilde{\varphi}}_{gi}(t^\prime),
\end{equation}
where $\omega_{ni} = 1/\sqrt{L_{ni} C_{ni}}$, and
\begin{equation}
G_{ni}(t) = -\frac{\hbar}{eL_{ni}} \left[ \frac{\sin (\omega_{ni} t)\,
    \dot{\varphi}_{ni}(0)}{\omega_{ni}} + \cos (\omega_{ni} t) \,
  \varphi_{ni}(0) \right].
\end{equation}
In addition, it is easy to show that the relation $\dot{\tilde{Q}}_i =
- \sum_{n=1}^\infty \dot{q}_{ni}$ also hold. Thus, we can write
\begin{equation}
\label{eq:tQ1}
\dot{\tilde{Q}}_i(t) + \frac{\hbar}{e} \int_0^t dt^\prime\,
Y_i(t-t^\prime)\, \dot{\tilde{\varphi}}_{gi}(t^\prime) = I_{Ni}(t)
\end{equation}
where the parameters $\{C_{ni}, L_{ni}\}$
must be chosen such that
\begin{equation}
Y_i(t) = \sum_{n=1}^\infty \frac{\cos(\omega_{ni} t)}{L_{n1}} \qquad
\rightarrow \qquad Y_i(\omega) = \frac{1}{Z_i(\omega)}
\end{equation}
and
\begin{equation}
I_{Ni}(t) = - \sum_{n=1}^\infty G_{ni}(t).
\end{equation}
%
%
\subsection{Voltage correlation functions}

We now turn the charge equation of motion, Eq. (\ref{eq:tQ1}), into
one for phase fluctuations, by using the following relationship:
\begin{equation}
\label{eq:varphig}
\dot{\tilde{\varphi}}_g = \frac{e}{\hbar} \left( V_g - V \right)
\end{equation}
and
\begin{equation}
\label{eq:VgQ}
V_g = \frac{Q}{\tilde{\cal C}} = \frac{\tilde{Q}}{\tilde{\cal C}} + V.
\end{equation}
Substituting Eq. (\ref{eq:VgQ}) into Eq. (\ref{eq:varphig}), we arrive at
\begin{equation}
\label{eq:dvarphi}
\dot{\tilde{\varphi}}_g = \frac{e}{\hbar}\,
\frac{\tilde{Q}}{\tilde{\cal C}},
\end{equation}
which allows us to retrieve the phase fluctuation equation-of-motion. 
\begin{eqnarray}
\label{eq:eqnmotionphi}
\tilde{\cal C} \cdot 
    \ddot{\tilde{\tilde{\varphi}}}_{g}(t)
    + \int_0^t
dt^\prime\, {\cal Y} (t-t^\prime) 
    \dot{\tilde{\tilde{\varphi}}}_{g}(t^\prime)=
\frac{e}{\hbar}\, {\cal I}_{N}(t).
\end{eqnarray}
Since we are interested in the behavior of a double-dot junction, with
each dot possessing its own charge and phase fluctuations, we will from
now on represent these quantities in a vector notation, as seen in
Eq. (\ref{eq:eqnmotionphi}). By applying a Fourier transformation and
substituting the random internal currents by external ones, we get
\begin{eqnarray}
i\omega\, {\cal Z}_t^{-1}(\omega) \cdot 
    \tilde{\tilde{\varphi}}_{g}(\omega)
     & = &
\frac{e}{\hbar}\,{\cal I}_{\rm pert\ }(\omega),
\end{eqnarray}
where
\begin{eqnarray}
{\cal Z}_t^{-1}(\omega) = {\cal Z}^{-1}(\omega) + i\omega\,
\tilde{\cal C}.
\end{eqnarray}
Now, substituting the external currents by appropriate generalized
force matrix,
\begin{eqnarray}
{\cal F}_{g}(\omega)  = \frac{\hbar}{e}\, {\cal I}_{\rm pert}(\omega) ,
\end{eqnarray}
we obtain
\begin{eqnarray}
\tilde{\tilde{\varphi}}_{g}(\omega) = {\cal
  X}_{\varphi_g} (\omega) \cdot {\cal F}_{g}(\omega),
\end{eqnarray}
where the dynamical susceptibility matrix is given by
\begin{eqnarray}
{\cal X}_{\varphi_g} (\omega) = \left( \frac{e}{\hbar} \right)^2
\frac{1}{i\omega} {\cal Z}_t(\omega),
\end{eqnarray}
whose imaginary part is given by
\begin{eqnarray}
{\cal X}^{\prime\prime}_{\varphi_g} (\omega) = -\left( \frac{e}{\hbar}
\right)^2 \, \frac{\mbox{Re} \left\{ {\cal Z}_t(\omega) \right\}}
  {\omega}.
\end{eqnarray}

Assuming that both transmission lines are at the same temperature, the
generalized form of the fluctuation-dissipation theorem reads
\begin{widetext}
\begin{equation}
\left( \begin{array}{cc} \left\langle \left| \tilde{\varphi}_{g1}
  (\omega) \right|^2 \right\rangle & \left\langle
  \tilde{\varphi}^\ast_{g1} (\omega)\, \tilde{\varphi}_{g2} (\omega)
  \right\rangle \\ \left\langle \tilde{\varphi}^\ast_{g2} (\omega)\,
  \tilde{\varphi}_{g1} (\omega) \right\rangle & \left\langle \left|
  \tilde{\varphi}_{g2} (\omega) \right|^2 \right\rangle \end{array}
\right) = \int_{-\infty}^\infty dt\, e^{-i\omega t}
\left( \begin{array}{cc} \left\langle \tilde{\varphi}_{g1} (t)\,
  \tilde{\varphi}_{g1} (0) \right\rangle & \left\langle
  \tilde{\varphi}_{g1} (t)\, \tilde{\varphi}_{g2} (0) \right\rangle
  \\ \left\langle \tilde{\varphi}_{g2} (t)\, \tilde{\varphi}_{g1} (0)
  \right\rangle & \left\langle \tilde{\varphi}_{g2} (t)\,
  \tilde{\varphi}_{g2} (0) \right\rangle \end{array} \right) =
\frac{-2\hbar}{1-e^{-\beta\hbar\omega}}\, {\cal
  X}^{\prime\prime}_{\varphi_g}(\omega).
\end{equation}
Hence,
\begin{equation}
\label{eq:varphicor}
\left( \begin{array}{cc} \left\langle \left| \tilde{\varphi}_{g1}
  (\omega) \right|^2 \right\rangle & \left\langle
  \tilde{\varphi}^\ast_{g1} (\omega)\, \tilde{\varphi}_{g2} (\omega)
  \right\rangle \\ \left\langle \tilde{\varphi}^\ast_{g2} (\omega)\,
  \tilde{\varphi}_{g1} (\omega) \right\rangle & \left\langle \left|
  \tilde{\varphi}_{g2} (\omega) \right|^2 \right\rangle \end{array}
\right) = \left( \frac{e}{\hbar} \right)^2 \frac{2\hbar}{\omega}\,
\frac{1}{1-e^{-\beta\hbar\omega}}\, \mbox{Re} \left\{ {\cal
  Z}_t(\omega)\right\}.
\end{equation}
\end{widetext}

We now turn to the fluctuations of the voltage at the dots. Since
\begin{equation}
\left( \begin{array}{c} \delta U_1 \\ \delta U_2 \end{array} \right) =
     {\cal C}^{-1} \cdot \left( \begin{array}{c} \delta \tilde{Q}_1
       \\ \delta \tilde{Q}_2 \end{array} \right) = \frac{\hbar}{e}\,
     {\cal C}^{-1} \cdot \tilde{\cal C} \cdot \left( \begin{array}{c}
       \dot{\tilde{\varphi}}_{g1}
       \\ \dot{\tilde{\varphi}}_{g2} \end{array} \right) ,
\end{equation}
where
\begin{equation}
\label{eq:C_cal}
{\cal C} = \left( \begin{array}{cc} C_1 & 0 \\ 0 & C_2 \end{array}
\right),
\end{equation} 
we find the matrix equation
\begin{equation}
\label{eq:correl}
{\cal U} = \frac{2\hbar\omega}{1-e^{-\beta\hbar\omega}}\, {\cal M},
\end{equation}
where
\begin{equation}
{\cal M} = {\cal C}^{-1} \cdot \tilde{\cal C} \cdot \mbox{Re} \left\{
{\cal Z}_t(\omega) \right\} \cdot \tilde{\cal C}^\dagger \cdot \left(
{\cal C}^{-1} \right)^\dagger,
\end{equation}
and
\begin{equation}
\label{eq:correlmat}
{\cal U} = \left( \begin{array}{cc} \left\langle \left| \delta U_1
  (\omega) \right|^2 \right\rangle & \left\langle \delta U^\ast_1
  (\omega)\, \delta U_2 (\omega) \right\rangle \\ \left\langle \delta
  U^\ast_2 (\omega)\, \delta U_1 (\omega) \right\rangle & \left\langle
  \left| \delta U_2 (\omega) \right|^2 \right\rangle \end{array}
\right).
\end{equation}
%

\section{Estimate of circuit parameters}
\label{sec:Estimates}

We now proceed to make realistic estimates of the effective circuit
parameters. The double-dot system is maintained at very low
temperatures, in the tens of millikelvin.\cite{kouwenhovenrev}
Typically, $k_B T \ll \Delta E$, $E_{C1}$, $E_{C2}$, where $\Delta E$
is the mean level spacing in the dots. The wires leading to the double
quantum dot are thermally anchored to a fridge at several temperature
stages (4~K, 1~K, 100~mK, and 10~mK). The transmission line resistance
$R_L$ is estimated to be 50~$\Omega$ for low temperatures (at or below
4~K) inside the dilution refrigerator, or 250~$\Omega$ in the copper
leads residing at room temperature. \cite{hanson}

The resistance of the two-dimensional electron gas (2DEG) can be
calculated using Drude's theory. \cite{ashcroft} The typical electron
density in a high-mobility GaAs 2DEG is approximately $n = 10^{11}$
cm$^{-2}$, which leads to an average Fermi velocity of about $v_F =
10^{5}$ m/s. At subkelvin temperatures, mean-free paths in the 2DEG
range from a few to up to one hundred $\mu$m.\cite{beenakker_rev}
Choosing $l = 10\ \mu$m, we arrive at a relaxation time $\tau = l/v_F
\approx 100$ ps, leading to an estimate of the low-temperature
conductivity of
\begin{equation}
\label{eq:sigma}
\sigma = \frac{n e^2 \tau}{m^\star} \nonumber \\ 
        \simeq  4.2 \times 10^{-2} \quad {\rm S},   
\end{equation}
with $m^\star = 0.067 m_e = 0.61 \times 10^{-31}$ kg being the
electron effective mass in GaAs. To calculate the resistance, we
considered a length $l = 10\ \mu$m and a width $w = 2.5\ \mu$m,
yielding a sheet resistance for the 2DEG underneath the gate
electrodes
\begin{equation}
\label{eq:R_elect}
R = \rho \frac{l}{w} \nonumber \\
  \simeq 95 \quad \Omega /\Box, 
\end{equation}
where $\rho = 1/\sigma$ is the resistivity of the 2DEG. This
resistance is responsible for a dissipative drag effect,
\cite{kycia01} that, for the sake of simplicity, will not be
considered in our model.

There is still one resistance left to be determined, which is the
resistance of the metallic electrodes. This resistance can be
determined by
\begin{equation}
\label{eq:R_elect2}
R = \rho \frac{l}{bc}, 
\end{equation}
where $\rho$ is the resistivity of the electrodes, approximately
$0.022 \times 10^{-8}\ \Omega$m for a Au electrode at low temperature
($<$ 4~K). If we consider the electrodes to have a $10\ \mu$m length
and a 30~nm $\times$ 60~nm cross section, we can estimate the
electrode resistance to be around 1~$\Omega$, a small value that will
also not be considered in our model.

The capacitance $C$ between the transmission line and the 2DEG was
estimated by solving the electromagnetic problem of a cylindrical
conducting wire of radius $r = 20$ nm placed at a distance of $d =
100$ nm from an infinite grounded conducting plate. Using the method
of images, we can estimate the total electric potential of this system
by integrating the electric field along the line connecting the
centers of the real and the image wires. This results in a capacitance
per unit length of $25$ aF/$\mu$m, and a total capacitance of $250$ aF
for a wire of $10\ \mu$m in length.

Any inductive couplings along our voltage lines can be estimated as
follows. For a metal electrode with rectangular cross section, the
self-inductance in ${\rm H/m}$ is approximated as \cite{engmanual}
\begin{equation}
L_{\rm rod} \sim 2l \left[
  \ln{\left(\frac{2l}{b+c}\right)}-\ln{\epsilon} + \frac{1}{2} \right]
\times 10^{-7}\, ,
\label{induct} 
\end{equation}
where $\epsilon$ is the aspect ratio of the electrode. For an
electrode with an aspect ratio of $2$, this equation yields $L \approx
1$ pH$/\mu$m. Thus, a $10\ \mu$m long electrode gives us an inductance
of $10$ pH. The parameters $C = 250$ aF and $L = 10$ pH, though useful
as rough estimates to characterize circuits, will not be used in our
model since they are very specific to the given circuit. In fact, in
order to estimate these circuit elements more precisely, more physical
parameters of the circuit in question would be necessary. To determine
the transmission line parameters in our model, we will make use of
Eqs. \eqref{TLwc} and \eqref{TLparam} from Sec. \ref{sec:transmline}
to give us a more general approach where we can model any transmission
line given these two operating parameters.  To give us a large enough
window to operate our qubits, we set our cutoff frequency to $\omega_c
= 200 \times 10^9$ rad/s. Table \ref{tbl:tl_parameters} summarizes the
transmission line parameters that fully describe $Z_i(\omega)$.

\begin{table}[!ht] 
\caption{Estimates for the transmission line parameters.}
\label{tbl:tl_parameters}
\begin{center}
\begin{tabular}{|l|c|} \hline 
  \multicolumn{2}{|c|}{Transmission line parameters} \\ \hline
Length $l$ & 10 $\mu$m \\
Transmission line capacitance $C_{t}$ & 10 pF \\
Transmission line inductance $L_{t}$ & 10 pH \\
Cutoff frequency $\omega_c$ & 200 $\times 10^9$ rad/s \\
$Z(\omega = 0) = R$ & 1 $\Omega$ \\ \hline
\end{tabular}
\end{center}
\vspace{-4ex}
\begin{minipage}{\textwidth}
\end{minipage}
\end{table}

The gate capacitance $C_{gi}$ $(i=1,2)$ for each quantum dot is given
by
\begin{equation}
C_{gi} = \frac{|e|}{\Delta V_{gi}}.
\label{GateCap}
\end{equation}
If we consider $\Delta V_{gi} \approx 4.5$ mV,\cite{dutch,
  kouwenhovenrev} we find $C_{gi} \approx 40$ aF.

Finally, we now estimate the tunneling parameters between the quantum
dots and the 2DEG. These are given by a tunneling junction with an
impedance $Z_T = R_T + jX_{C_{T}}$. We can obtain a lower bound for
the tunneling resistance $R_T$ by estimating the inverse of the
Coulomb blockade peak conductance. In the regime $\Gamma \ll k_B T$,
$G_{\rm max}$ is given by \cite{beenakker}
\begin{equation}
G_{\rm max} = \frac{e^2}{4 k_B T}\frac{\Gamma^l \Gamma^r}{\Gamma^l +
  \Gamma^r},
\end{equation}
where tunneling rates of an electron through the potential barrier
into (or out of) each dot are assumed equal for the sake of simplicity
($\Gamma^l = \Gamma^r$) For an electron temperature in the dot $T
\approx 150$ mK and a peak conductance height of $2 \times 10^{-3}
e^2/h$, \cite{petta04} we find the tunneling resistance to be larger
than or on the order of $10$ M$\Omega$.
We can estimate the tunneling capacitance indirectly. We know the
expression for the total capacitance of a flat disk to be
\begin{equation}
C_i = 8 \epsilon_r \epsilon_0 R.
\end{equation}
Assuming $R\simeq 80$ nm as the radius of the quantum dot and
$\epsilon_r \approx 11$ for GaAs at high frequencies, yielding a total
capacitance $C_i \approx 60$ aF for each quantum dot.

From the total capacitance we can estimate the interdot capacitance
between dots 1 and 2, since
\begin{equation}
C_{m} = \frac{\Delta V^m_{gi}}{\Delta V_{gi}} C_j,
\label{InterdotCap}
\end{equation}
where $i \neq j$. For $\Delta V^m_{gi} \approx 0.4$ mV, \cite{dutch,
  kouwenhovenrev} we find $C_{m} \approx 5$ aF.

The total capacitance for each quantum dot, as seen previously, is the
sum of all capacitances attached to the dot. As such, by knowing $C_m
= 5$ aF and $C_{Ti} = 40$ aF, we find $C_{gi}$ $\approx 15$ aF.

Using these estimates for the circuit elements, we are able to
determine the distributed parameters of our noisy transmission
lines. According to Eq. \eqref{TLparam}, if we assume a cutoff
frequency of $\omega_c \sim 10^{11}$ Hz, we find $L_{ti} \sim 1$
pH/$\mu$m and $C_{ti} \sim 1$ pF/$\mu$m.

In principle, one can also consider the ground (2DEG) to be a source
of noise, and as such it can also be modeled by means of a
frequency-dependent impedance. This would require however an
appropriate estimate of the inductance along the 2DEG. We did not
carry out such an estimate. However, we attempt to take into account
the coupling between the quantum dot leads. This coupling is given by
the lumped capacitance $C_{12}$, as shown in
Fig. \ref{fig:DQDcirc}. This capacitance was estimated to be
approximately $20$ aF by means of numerical multipole expansion
calculations performed by a field solver software.\cite{fastcap} We
summarize in Table \ref{tbl:circuit_parameters} the relevant circuit
parameters necessary to fully characterize the DQD setup.
\begin{table}[!ht] 
\caption{Estimates for the DQD circuit parameters. $i=1,2$,
  corresponding to each quantum dot}
\label{tbl:circuit_parameters}
\begin{center}
\begin{tabular}{|l|c|} \hline 
  \multicolumn{2}{|c|}{Circuit parameters} \\ \hline Transmission line
  capacitance $C_{ti}$ & 1 pF/$\mu$m \\ Transmission line inductance
  $L_{ti}$ & 1 pH/$\mu$m \\
Interdot capacitance $C_m$ & 5 aF \\ Tunneling capacitance $C_{Ti}$ &
40 aF \\ Tunneling resistance $R_{Ti}$ & $\gtrapprox$ 10 M$\Omega$
\\ Gate capacitance $C_{gi}$ & 15 aF \\ Total quantum dot capacitance
$C_i$ & 60 aF \\ Capacitive coupling between transmission lines
$C_{12}$ & $\simeq$ 20 aF \\ Electrode resistance $R_{i}$ & 1 $\Omega$
\\ \hline
\end{tabular}
\end{center}
\vspace{-1ex}
\end{table}

\section{Bounds on decoherence rates and $Q$ factors}
\label{sec:rates}

Through the fluctuation-dissipation theorem, we can relate the
impedance $Z_{1,2}(\omega)$ to a source of electromagnetic gate
fluctuations $\delta V_{g1,2}$. These gate fluctuations $\delta V_{gi}
= Q_{0,i}/C_{gi} (i=1,2)$ can be determined through the
diagonalization of the Hamiltonian in Eq. \eqref{hamiltTL}. We
consider in this paper the case of Johnson-Nyquist noise.
\cite{weiss,makhlin03} Following the standard procedure, we relate the
energy relaxation rate to the power spectrum of voltage fluctuations,
\begin{equation}
\gamma_1 = \frac{\sin^2\eta}{4\hbar^2}\, \left[ \frac{S_{\delta
      U}(\Delta/\hbar) + S_{\delta U}(-\Delta/\hbar)}{2} \right],
\end{equation}
where
\begin{equation}
S_{\delta U}(\omega) = \frac{e^2}{2} \left\langle \left| \left[ \delta
  U_1(\omega) - \delta U_2(\omega) \right] \right|^2 \right\rangle.
\end{equation}
and $\tan \eta = \Delta/\epsilon$, see Eq. (\ref{eq:correl}). Using
Eq. (\ref{eq:Hs}), we obtain
\begin{equation}
\gamma_1 = \frac{\sin^2\eta\, \Delta}{\hbar\, R_K} \coth\left(
\frac{\Delta}{2 k_B T} \right)\, {\tilde {\cal M}}
\label{relrate}      
\end{equation}
where
\begin{eqnarray}
{\tilde {\cal M}} & = & {\cal M}_{11} + {\cal M}_{22} - {\cal M}_{12}
- {\cal M}_{21}.
\label{u_tilde}      
\end{eqnarray}
and $R_K$ is the resistance quantum ($ = h/e^2 \simeq 25.8 {\rm
  k}\Omega $). From this expression we can calculate the energy
relaxation and dephasing times,
\begin{equation}
T_1 = 1 /\gamma_1 = \frac{\hbar\, R_K}{\sin^2\eta\, \Delta}
\frac{\tanh(\Delta /2 k_B T)}{ {\tilde {\cal M}}},
\label{reltime}      
\end{equation}
and
\begin{equation}
T_2 = \left[ \frac{1}{2 T_1} +
  \frac{\cos^2\eta}{4\hbar^2}\ S(\Delta/\hbar) \right]^{-1}.
\label{dephtime}      
\end{equation}
Hereafter, we will assume zero bias ($\epsilon=0$), in which case
$\eta=\pi/2$ and $T_2 = 2T_1$. The quality factor of the quantum
oscillations is then given by
\begin{equation}
Q = \omega_{\rm osc}\, T_2 = \frac{\omega_{\rm osc}}{\pi \gamma_1}
\label{qfactor}      
\end{equation}
where $\omega_{\rm osc}$ is the frequency of quantum oscillations
observed in the DQD system, as defined by \cite{weiss}
\begin{equation}
\omega_{\rm osc} = \sqrt{ 2\frac{\Delta}{\hbar} \left(
  2\frac{\Delta}{\hbar} + \frac{\gamma_2}{2} \right) -
  \frac{\gamma^2_1}{4}},
\label{oscfreq}      
\end{equation} 
with $\Delta$ being the potential barrier height between quantum dots,
as shown in Eq. \eqref{eq:Hs}, and $\gamma_2$ being defined as
\begin{equation}
\gamma_2 = - \dashint_{0}^{\infty} \frac{dy}{y^2-1}\, \nu(2\Delta y)\,
      {\rm coth} \left( \frac{\Delta y}{k_B T} \right),
\label{eq:gamma2}      
\end{equation} 
where $\nu$ is the bath spectral function, defined as
\begin{eqnarray}
\nu(\omega) & = & \frac{2}{\pi} \frac{\omega}{R_K} \Big\{ \left\langle
\left| \delta U_1 (\omega) \right|^2 \right\rangle + \left\langle
\left| \delta U_2 (\omega) \right|^2 \right\rangle \nonumber \\ & & -
\left\langle \delta U^\ast_1 (\omega)\, \delta U_2 (\omega)
\right\rangle - \left\langle \delta U^\ast_2 (\omega)\, \delta U_1
(\omega) \right\rangle \Big\}.
\label{eq:specfunction}      
\end{eqnarray}

The operating frequency $\omega = 2\Delta /\hbar$ is fed to the
circuit by the voltage generators and carried through the gates.  The
other terms in Eq. \eqref{oscfreq}, as it turns out, are small enough
corrections to the operating frequency so that they may be ignored.
Thus, from now on we will assume $\omega_{\rm osc} = \omega$.

We will now analyze in detail two different scenarios: one where the
transmission lines are decoupled, while the other includes the
capacitive coupling $C_{12}$ between transmission lines, as seen in
Fig. \ref{fig:DQDcirc}.

\subsection{Case (i): Decoupled transmission lines}
\label{sec:decouplTL}   

\begin{figure}[ht]
\centering
\includegraphics[width=8cm]{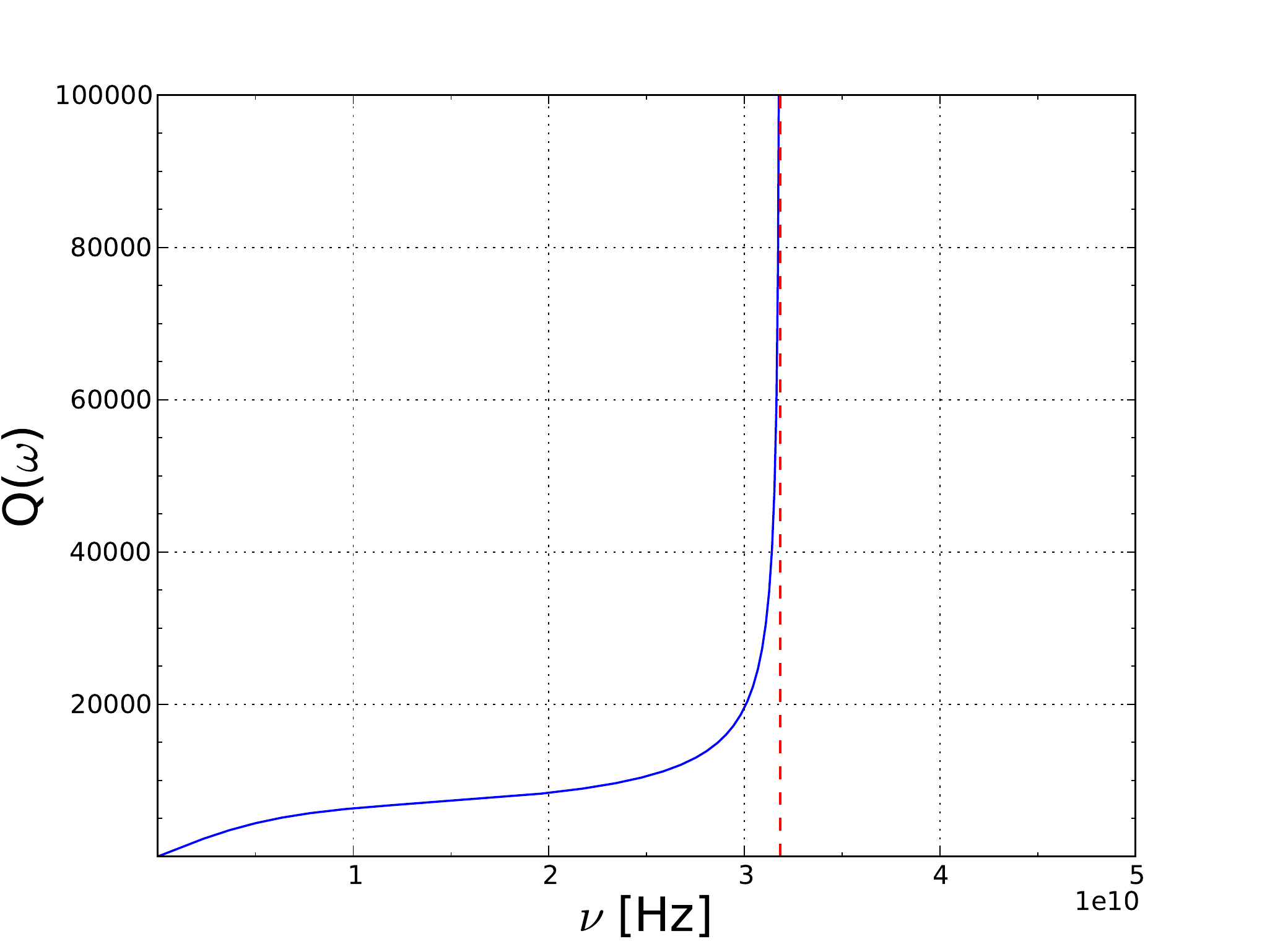}
\caption{Qubit quality factor as a function of frequency for two
  decoupled semi-infinite transmission lines, with temperature $T=
  150$ mK and the circuit parameters presented in Table
  \ref{tbl:circuit_parameters}.}
\label{fig:Q_dec_long}
\end{figure}

\begin{figure}[ht]
\centering
\includegraphics[width=8cm]{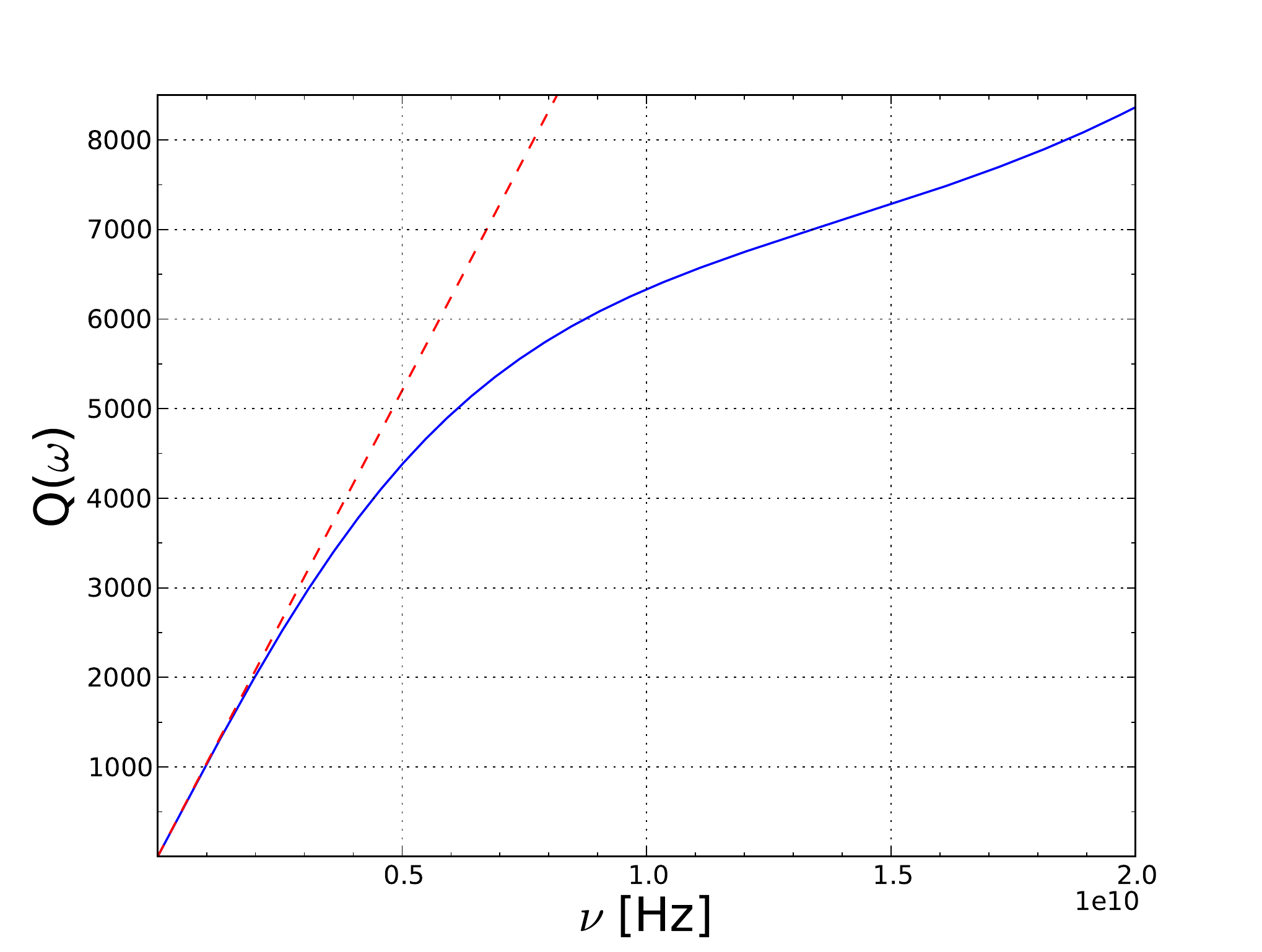}
\caption{Qubit quality factor as a function of frequency for $\nu <
  20$ GHz and two decoupled semi-infinite transmission lines with the
  same parameter values as in Fig. \ref{fig:Q_dec_long}.}
\label{fig:Q_dec}
\end{figure}

It is useful to look at the case where there is no coupling between
the electrodes. The decoherence introduced by the electromagnetic
voltage fluctuations can still be analyzed using
Eqs.~\eqref{relrate}--\eqref{qfactor}, but some simplifications to
the impedance matrix are now possible. This case corresponds to having
$C_{12} = 0$, so the matrix ${\cal Z}$ from Eq. \eqref{eq:Z_cal} is
reduced to
\begin{equation}
\label{eq:Z_cal_dec}
{\cal Z} = \left( \begin{array}{cc} Z_1 \left( 1 + \frac{C_m}{C_1}
  \right) & -Z_1\frac{C_m}{C_2} \\ -Z_2\frac{C_m}{C_1} & Z_2 \left( 1
  + \frac{C_m}{C_2} \right) \end{array} \right).
\end{equation}
In this case we observe the highest possible quality factors for our
double-quantum-dot setup, as seen in Figs.~\ref{fig:Q_dec_long}, and
~\ref{fig:Q_dec}.

\begin{figure}[ht]
\centering
\includegraphics[width=8cm]{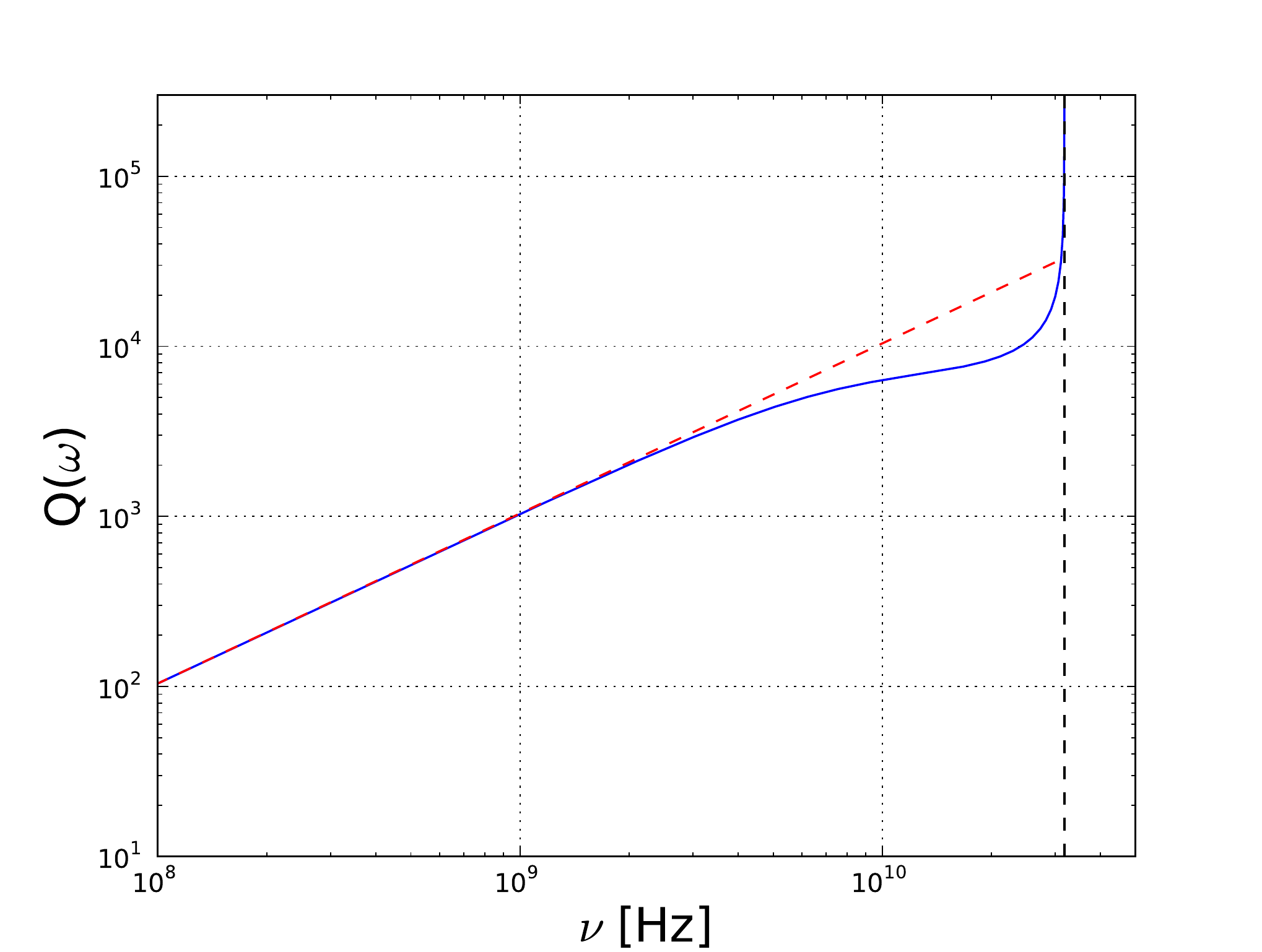}
\caption{Quality factor as a function of frequency for two decoupled
  transmission lines represented in a logarithmic scale with the same
  parameter values as in Fig. \ref{fig:Q_dec_long}.}
\label{fig:Q_declog_long}
\end{figure}

If we look back at Eq. \eqref{qfactor} and take its asymptotic limit
for low frequencies, Eq. \eqref{eq:Z_cal_dec} is then reduced to
\begin{equation}
\label{eq:Z_cal_asy}
{\cal Z} = \left( \begin{array}{cc} R_1 \left( 1 + \frac{C_m}{C_1}
  \right) & -R_1\frac{C_m}{C_2} \\ -R_2\frac{C_m}{C_1} & R_2 \left( 1
  + \frac{C_m}{C_2} \right) \end{array} \right),
\end{equation}
where $R_i = {\cal Z}(\omega = 0) = \sqrt{L_{ti}/C_{ti}}$, as reported
earlier, and with the assumption that $C_1 = C_2$, $C_{g1} = C_{g2}$,
and $R_1 = R_2$. This, combined with the fact that $\mbox{Re} \left\{
{\cal Z}_t(\omega) \right\} \to \mbox{Re} \left\{ {\cal Z}(\omega)
\right\}$ for $\omega \to 0$, yields
\begin{equation}
\lim_{\omega \to 0} Q(\omega)  =  (8.9 \times 10^{-7} [s])\, \nu , 
\label{Q_dec_asymp}      
\end{equation}
where we notice a linear dependence of $Q$ with respect to $\nu$, as
can also be evidenced in the log-log graph shown in
Fig. \ref{fig:Q_declog_long}. While $R_i$ is an important modeling
parameter for the transmission lines, it is also clear that $C_{ti}$
and $L_{ti}$ ultimately influence how quickly this linear regime
establishes itself once we move to lower frequencies.

\begin{figure}[ht]
\centering
\includegraphics[width=8cm]{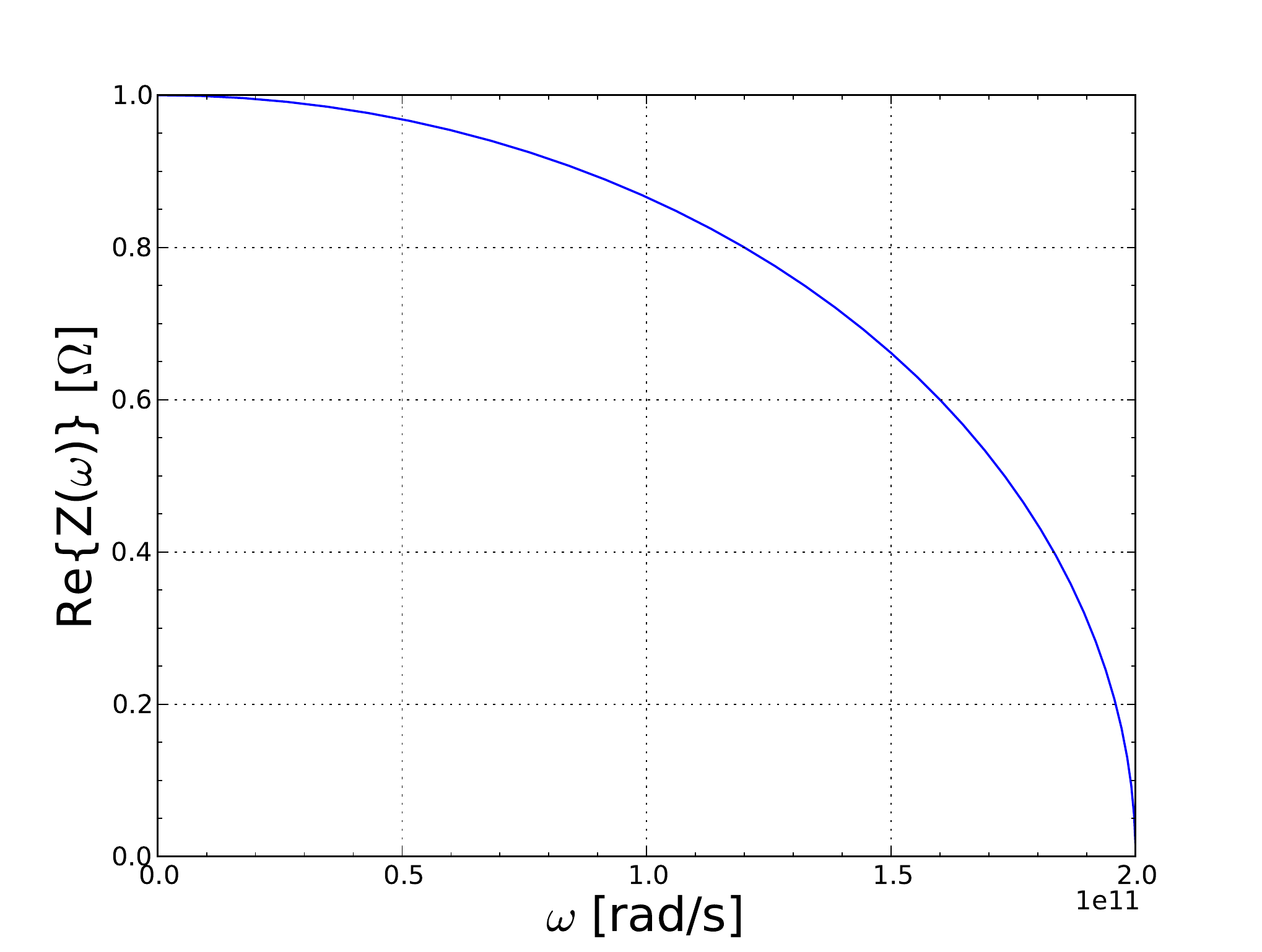}
\caption{Real part of the impedance ${\cal Z}(\omega)$ as a function
  of the frequency $\omega$. Transmission line parameters are defined
  in Table \ref{tbl:circuit_parameters}.}
\label{fig:Z_w}
\end{figure}

Turning our attention now to higher frequencies, we notice an
important characteristic of the transmission lines.  The real part of
the transmission line impedance $\mbox{Re} \left\{ {\cal Z}(\omega)
\right\}$ has a cutoff frequency given by $\nu_c = \omega_c / 2 \pi$.
In Fig.~\ref{fig:Z_w}, it can be seen that as $\omega \to \omega_c$,
$\mbox{Re} \left\{ {\cal Z}(\omega) \right\} \to 0$, making $\mbox{Re}
\left\{ {\cal Z}_t(\omega) \right\} \to 0$ as well, causing the
quality factor $Q$ to diverge at $\omega = \omega_c$.

\subsection{Case (ii): Capacitively coupled transmission lines}
\label{sec:capcouplTL}  

\begin{figure}[ht]
\centering
\includegraphics[width=8cm]{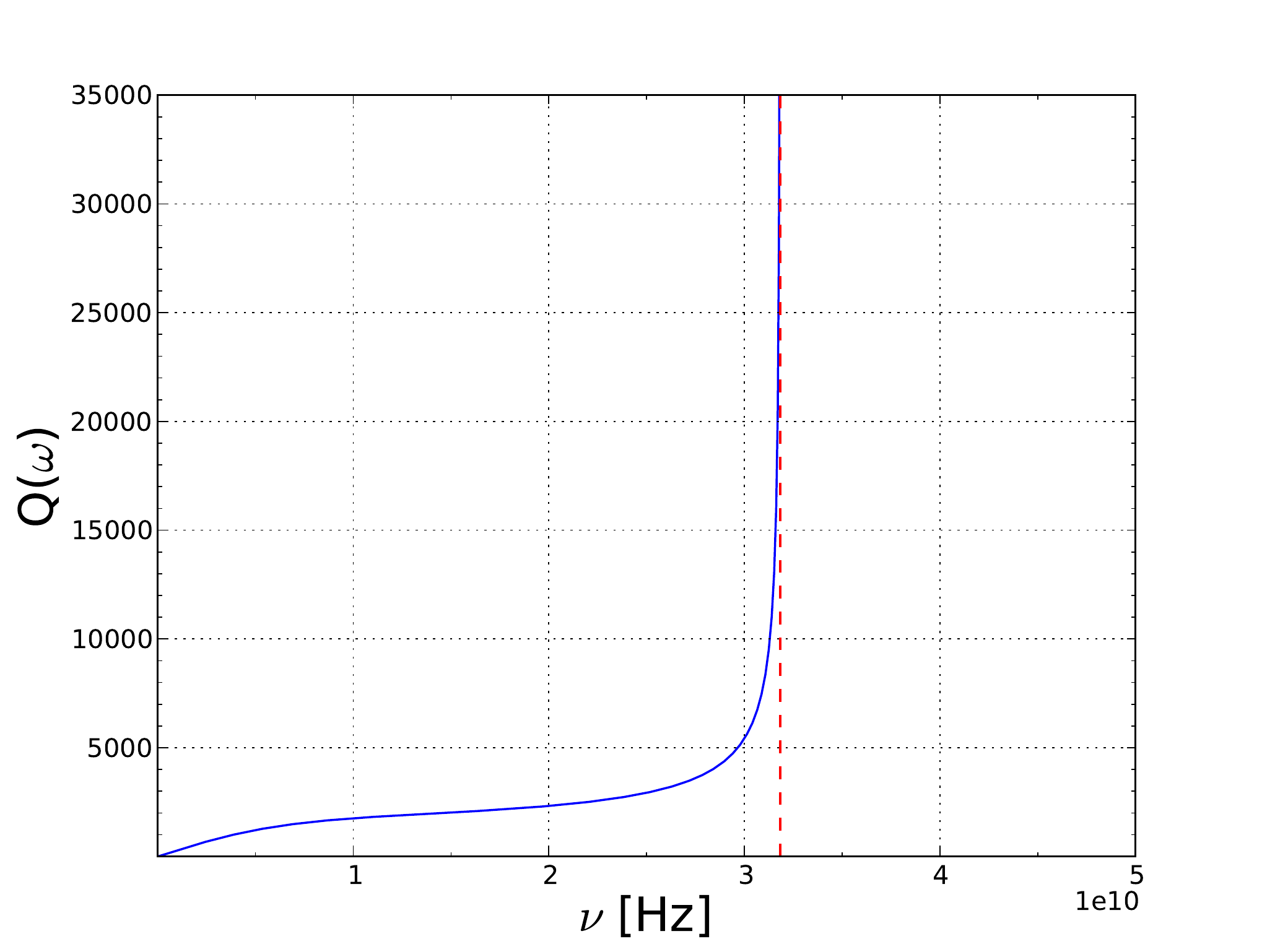}
\caption{Qubit quality factor as a function of frequency, with
  temperature $T = 150$ mK. The circuit parameters utilized are
  presented in Table \ref{tbl:circuit_parameters}.}
\label{fig:Q_w}
\end{figure}

\begin{figure}[ht]
\centering
\includegraphics[width=8cm]{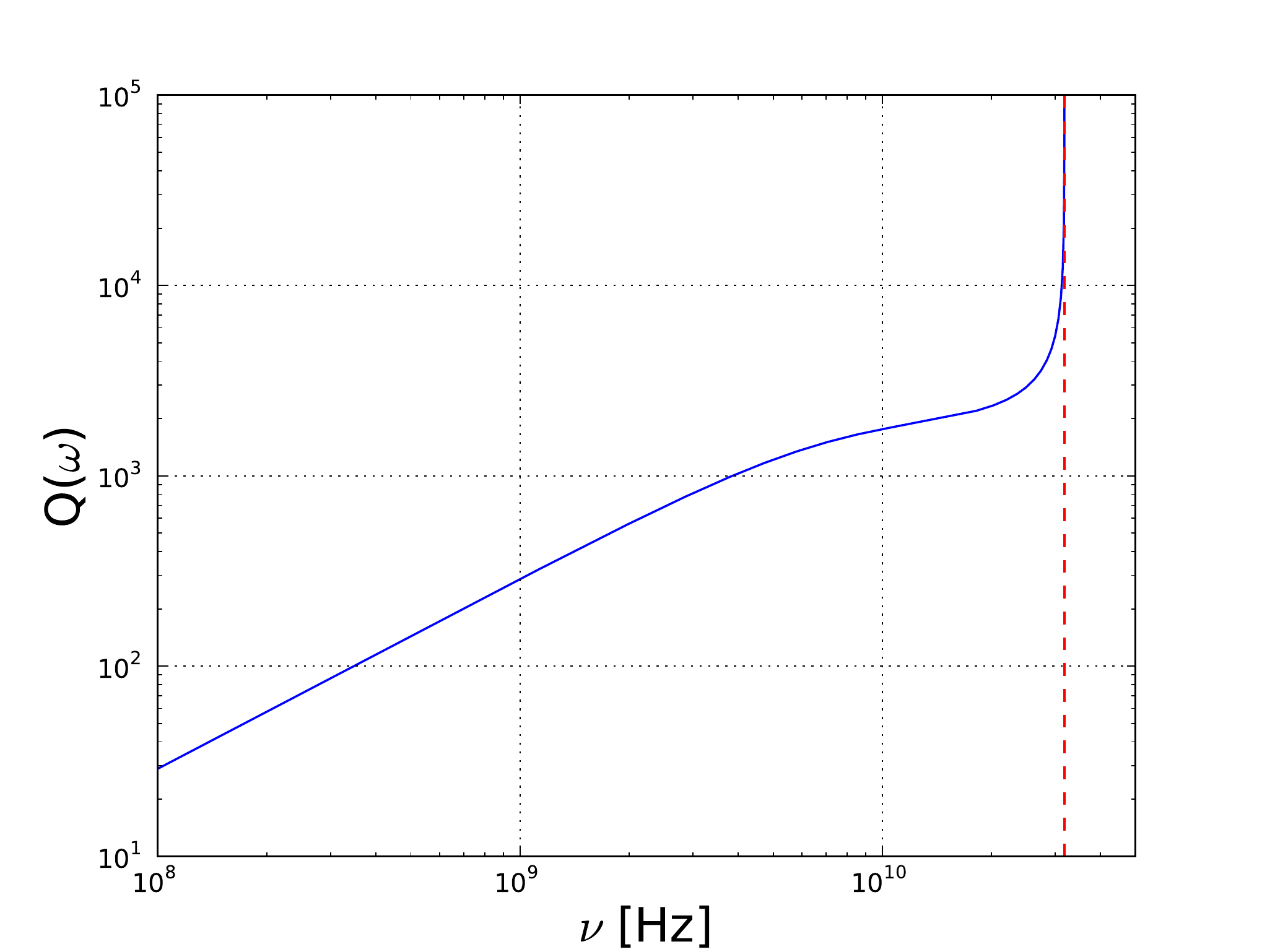}
\caption{Qubit quality factor as a function of frequency represented
  in a logarithmic scale. The circuit parameters utilized are the same
  as in Fig. \ref{fig:Q_w}.}
\label{fig:log_Q_w}
\end{figure}

\begin{figure}[ht]
\centering
\includegraphics[width=8cm]{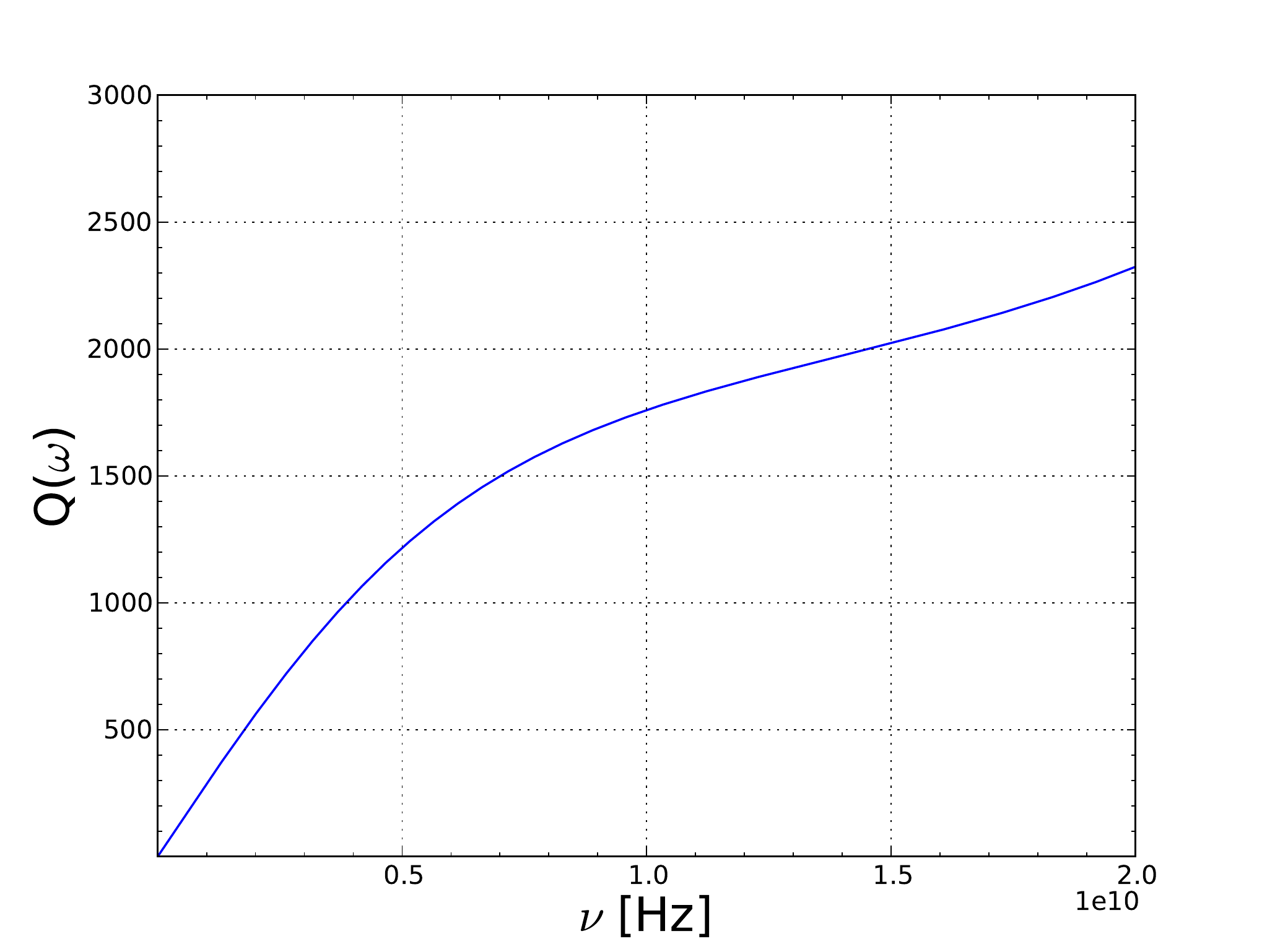}
\caption{Quality factor as a function of frequency for $\nu < 20$ GHz
  and two decoupled semi-infinite transmission lines with the same
  parameter values as in Fig. \ref{fig:Q_w}.}
\label{fig:Q_w_zoomed}
\end{figure}

Inserting now the inter-capacitive coupling $C_{12}$ estimated in
Sec.~\ref{sec:Estimates}, we obtain the quality factor $Q$ as a
function of frequency $\nu$ shown in Figs.~\ref{fig:Q_w},
~\ref{fig:log_Q_w}, and ~\ref{fig:Q_w_zoomed}. In Fig.~\ref{fig:Q_w},
we can clearly observe the quality factor diverge at the frequency
$\nu_c \simeq $ 320 GHz, corresponding to the cutoff frequency. From
now on we shall restrict our discussion to operating frequencies under
20 GHz (Fig.~\ref{fig:Q_w_zoomed}), which are more realistic for
practical implementations of qubit operations.

\begin{figure}[ht]
\centering
\includegraphics[width=8cm]{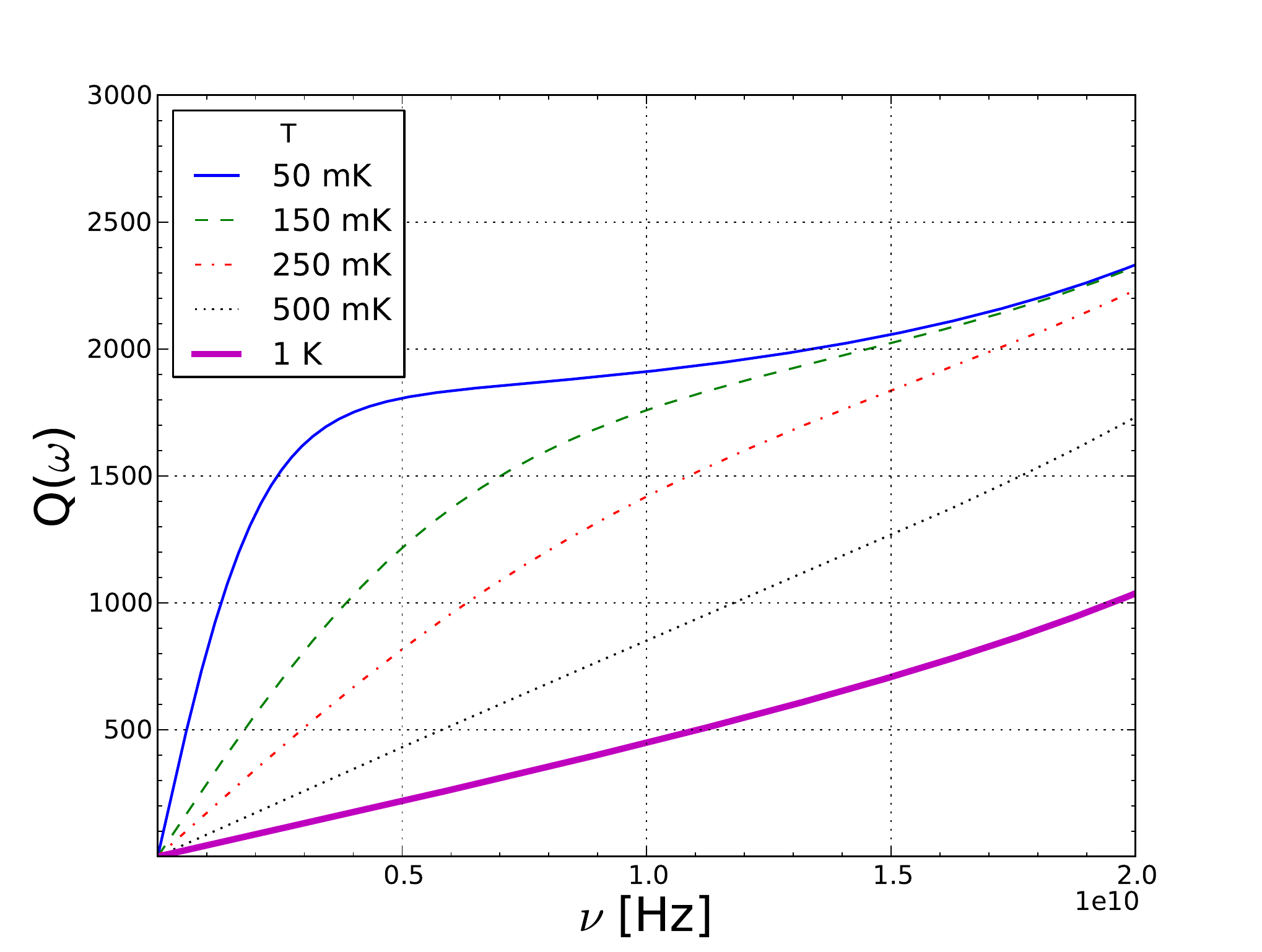}
\caption{Qubit quality factor as a function of operating frequency for
  temperatures $T= 50, 150, 250, 500$ mK, and 1 K. The circuit
  parameters utilized are presented in Table
  \ref{tbl:circuit_parameters}.}
\label{fig:Q_Ts}
\end{figure}

\begin{figure}[ht]
\centering
\includegraphics[width=8cm]{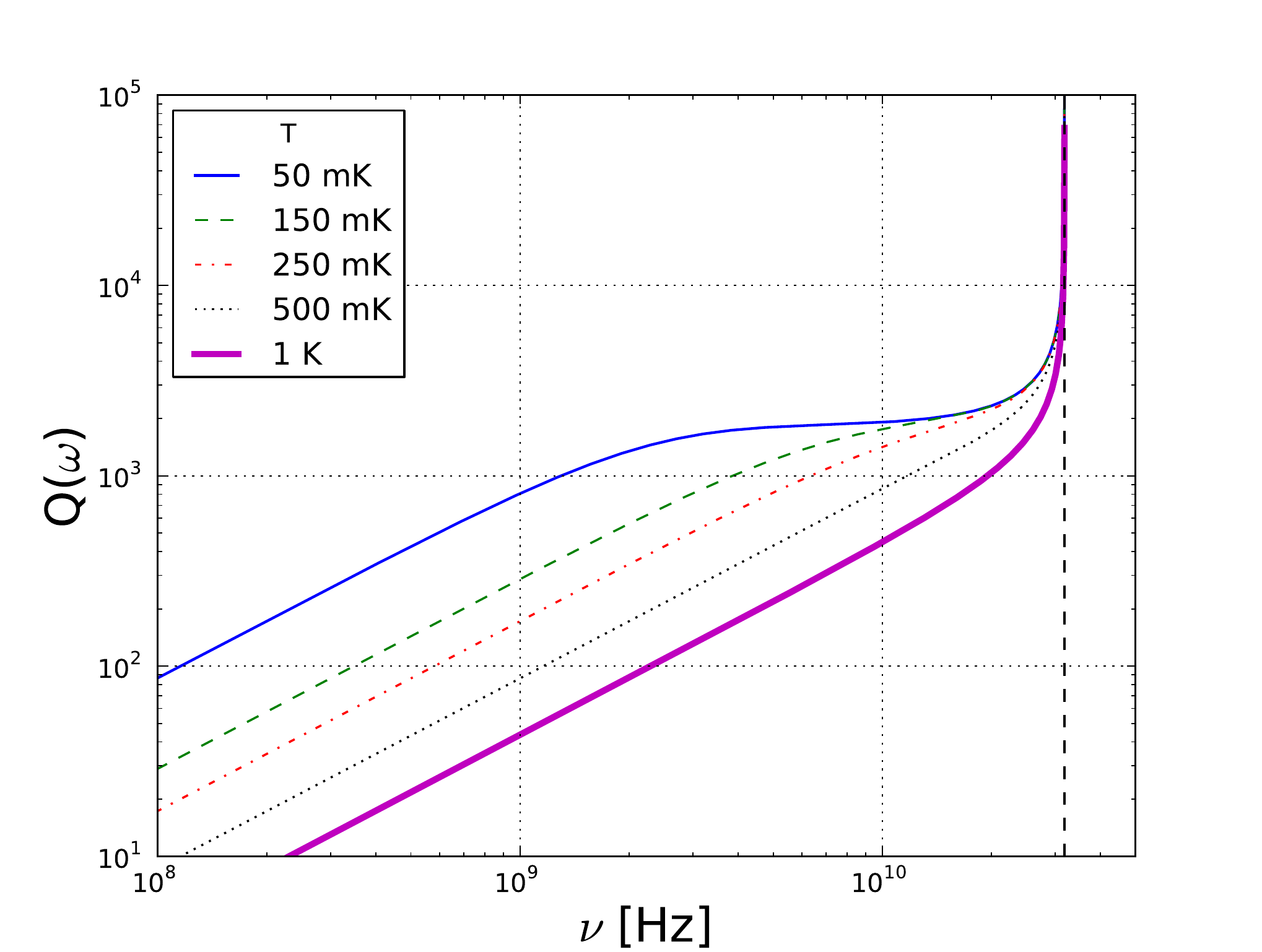}
\caption{Logarithmic representation of the qubit quality factor as a
  function of operating frequency for temperatures $T = 50, 150, 250,
  500$ mK, and 1 K. The circuit parameters utilized are presented in
  Table \ref{tbl:circuit_parameters}.}
\label{fig:log_Q_Ts}
\end{figure}

It is interesting to observe the influence of temperature on the
decoherence introduced into the system by voltage fluctuations. We
show below, in Figs. \ref{fig:Q_Ts} and \ref{fig:log_Q_Ts}, a family
of $Q$ factor curves as a function of operating frequency $\nu$ for
temperatures ranging from $50$ mK all the way to room temperature.  As
temperature increases, more environmental modes are available for the
system to couple with, effectively increasing dissipative effects.

\begin{figure}[ht]
\centering
\includegraphics[width=8cm]{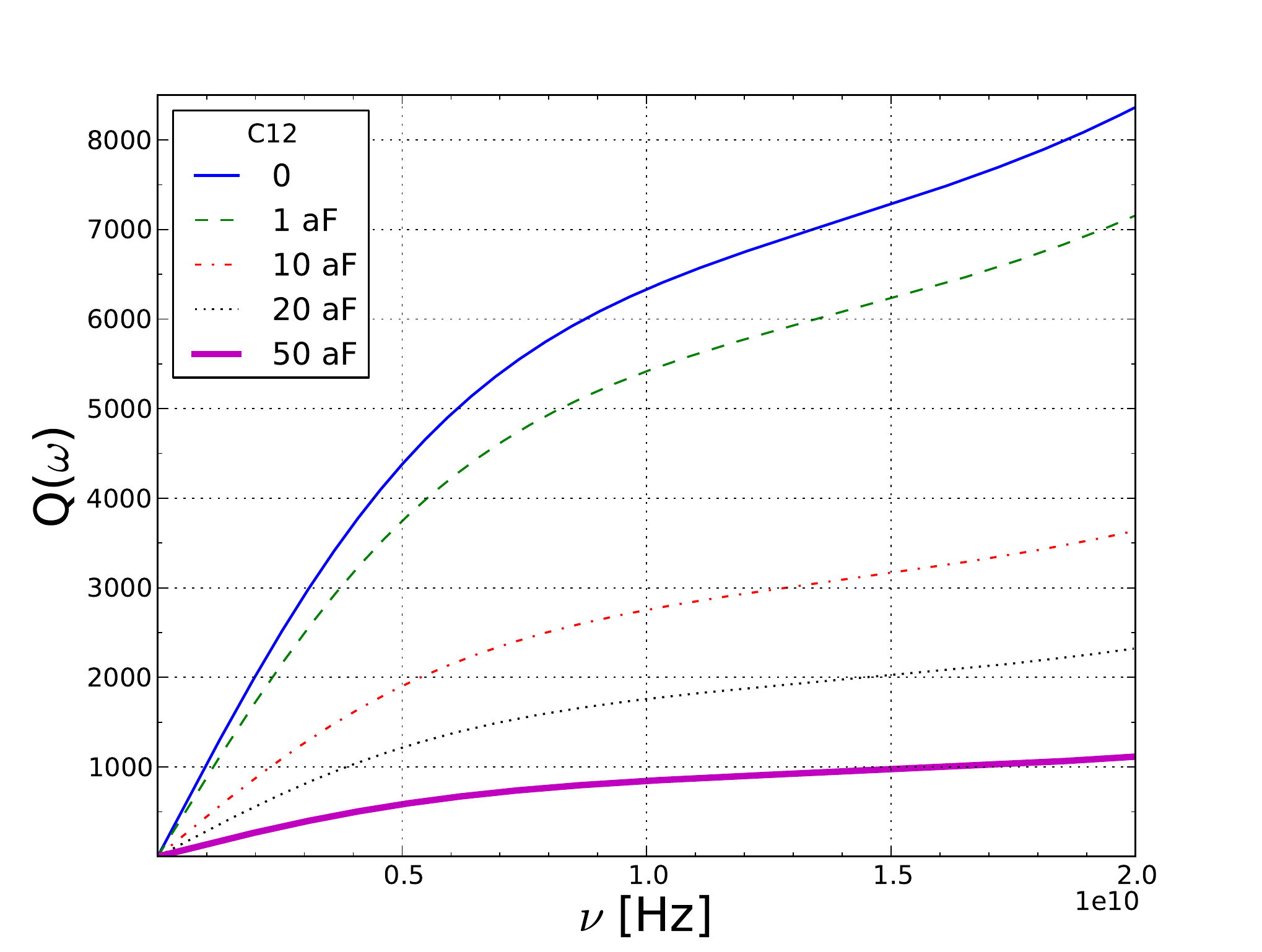}
\caption{Quality factor as a function of operating frequency for
  temperature $T = 150$ mK and inter-capacitive couplings $C_{12} = $
  0, 1.3 , 10, 20, and 50 aF. The circuit parameters utilized are
  presented in Table \ref{tbl:circuit_parameters}.}
\label{fig:Q_C12s}
\end{figure}

\begin{figure}[ht]
\centering
\includegraphics[width=8cm]{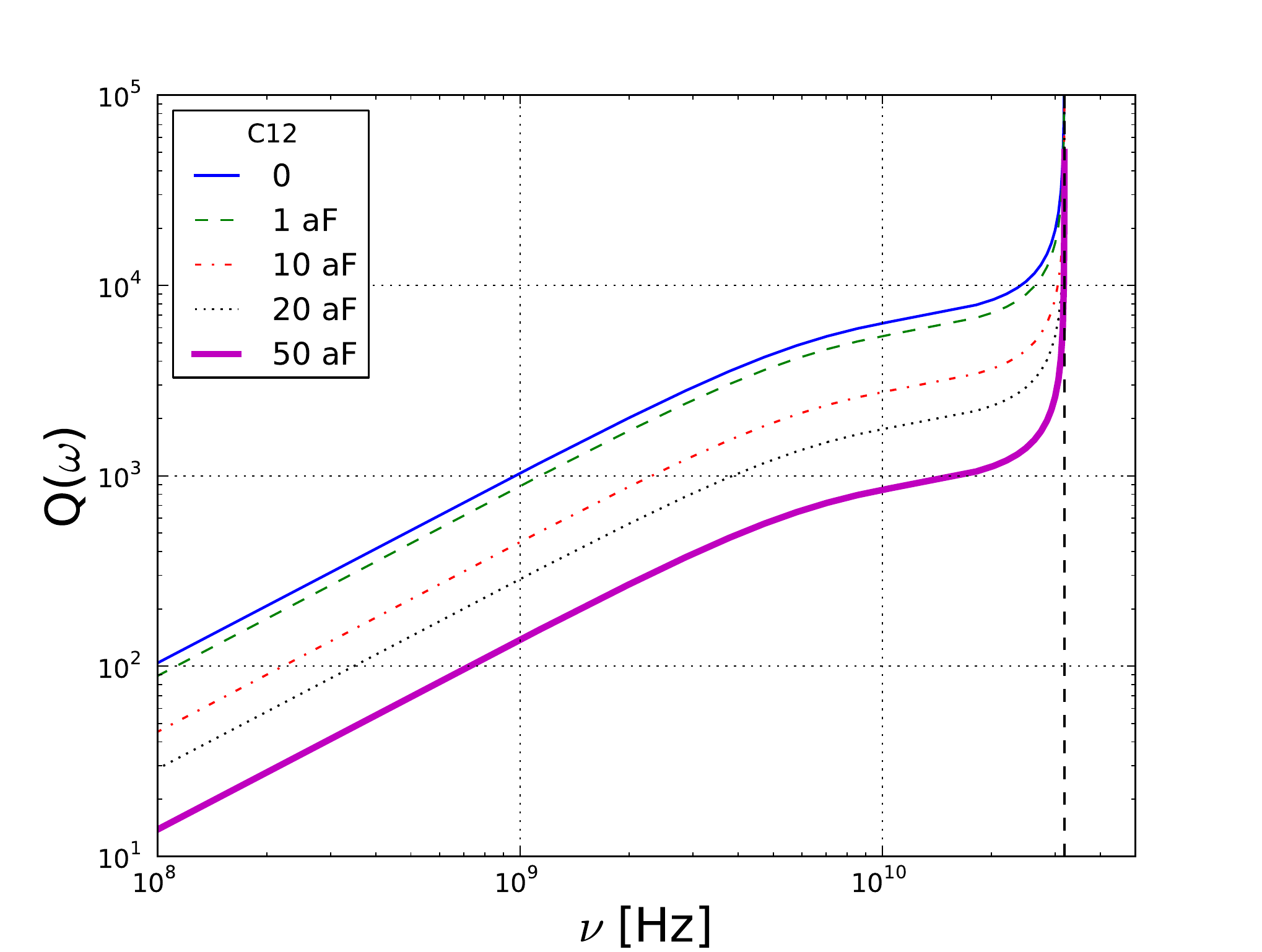}
\caption{Logarithmic representation of the qubit factor as a function
  of operating frequency for temperature $T = 150$ mK and
  inter-capacitive couplings $C_{12} = $ 0, 1.3, 10, 20, and 50
  aF. The circuit parameters utilized are the same as in
  Fig. \ref{fig:Q_C12s}.}
\label{fig:log_Q_C12s}
\end{figure}

We can also observe the influence of the inter-capacitive coupling
$C_{12}$ on the quality factor, as seen in Figs. \ref{fig:Q_C12s} and
\ref{fig:log_Q_C12s}. For weaker coupling, i.e., smaller $C_{12}$, the
quality factors are higher, as $C_{12}$ approaches the limiting case
of decoupled lines. Note that $Q(\omega)$ will still not reach the
same levels of the decoupled case due to the presence of the
capacitance $C_m$.

In Tables \ref{tbl:deco_T2} and \ref{tbl:deco_Q} we present the
results of calculations for the decoherence time $T_2$ and the $Q$
factor for several different values of temperature $T$ and
inter-capacitive coupling $C_{12}$. It is easy to understand why
higher temperatures degrade decoherence times in qubit operations. We
can consider two extreme cases, namely, one where the electrical leads
are inside a dilution refrigerator and another where they are at room
temperature. We will also consider an operating frequency $\nu =
\omega /2 \pi$ of $10$ GHz. First, let us assume that leads connected
to the gate electrode are inside the dilution refrigerator. In this
case, a temperature $T = 150$ mK results in a relaxation time $T_1 =
88$ ns and a decoherence time of $T_2 = 176$ ns. This scenario yields
a quality factor of $Q \approx 1760$. If we consider now the case
where the leads are at room temperature, we estimate the relaxation
time and the dephasing time to be approximately $76$ ps and $152$ ps,
respectively, resulting in a quality factor of $Q \simeq 1.5$, more
than 1000 times lower.
\begin{table}[ht] 
\caption{Estimates for the dephasing times $T_2$ for different values
  of temperature $T$ and interline capacitive coupling $C_{12}$.}
\label{tbl:deco_T2}
\begin{center}
\begin{tabular}{|c||c|c|c|c|c|} \hline 
  \multicolumn{6}{|c|}{Dephasing time $T_{2}$ (ns)} \\ \hline
{T (K)} & \multicolumn{5}{|c|}{$C_{12}$ (aF)} \\ &
\multicolumn{1}{c}{0} & \multicolumn{1}{c}{$\sim$1} &
\multicolumn{1}{c}{10} & \multicolumn{1}{c}{20} &
\multicolumn{1}{c|}{50} \\ \hline
$50 \times 10^{-3}$ & 688 & 588 & 300 & 191 & 92 \\ 
$150 \times 10^{-3}$ & 633 & 542 &  275 & 176 & 84 \\ 
$250 \times 10^{-3}$ & 511 & 437 & 222 & 142 & 68 \\ 
$500 \times 10^{-3}$ & 306 & 262 & 133 & 85 & 41 \\ 
$1$ & 161 & 138 & 70 & 45 & 22 \\ 
$300$ & 0.55 & 0.47 & 0.24 & 0.15 & 0.07 \\ \hline
\end{tabular}
\end{center}
\end{table}
A much more interesting analysis stems from varying the
inter-capacitive coupling between the transmission lines. For higher
values of $C_{12}$, it would be intuitive to expect both transmission
lines to be more strongly coupled, meaning that decoherence in the
system would be weaker since voltage fluctuations in the two lines
would be correlated. As it turns out, however, the stronger coupling
between transmission lines results in larger off-diagonal terms in the
matrix of voltage correlations defined in Eq. \eqref{eq:correl}. If we
look at Eq. \eqref{qfactor} once more, it is easy to see that larger
off-diagonal terms subtracted from the main diagonal correlation terms
results in smaller $Q$ factors, as evidenced by the behavior of the
family of $Q$ factor curves in Fig.~\ref{fig:Q_C12s} for different
values of inter-capacitive coupling and the calculated values
presented in Table ~\ref{tbl:deco_Q}.
\begin{table}[!ht] 
\caption{Estimates of $Q$ factors for different values of temperature
  $T$ and inter-capacitive coupling $C_{12}$.}
\label{tbl:deco_Q}
\begin{center}
\begin{tabular}{|c||c|c|c|c|c|} \hline 
  \multicolumn{6}{|c|}{$Q$ factor} \\ \hline
  \multicolumn{1}{c}{0} & \multicolumn{1}{c}{$\sim$1} &
  \multicolumn{1}{c}{10} & \multicolumn{1}{c}{20} &
  \multicolumn{1}{c|}{50} \\ \hline
$50 \times 10^{-3}$ & 6878 & 5884 & 2990 & 1910 & 917 \\ $150 \times
  10^{-3}$ & 6333 & 5418 & 2753 & 1760 & 844 \\ $250 \times 10^{-3}$ &
  5108 & 4369 & 2220 & 1418 & 681 \\ $500 \times 10^{-3}$ & 3059 &
  2617 & 1329 & 850 & 408 \\ $1$ & 1614 & 1380 & 702 & 448 & 215
  \\ $300$ & 5.5 & 4.7 & 2.4 & 1.5 & 0.7 \\ \hline
\end{tabular}
\end{center}
\end{table}

\section{Electromagnetic fluctuations in double-dot spin qubits}
\label{sec:spinqubits}

Decoherence due to the coupling between orbital (charge) and
environmental degrees of freedom also occurs in certain spin-based
quantum dot qubits. For instance, in the double-dot system introduced
by the Harvard group \cite{petta05}, the computational basis is formed
by the singlet and the $S_z=0$ triplet states of a DQD system
possessing an overall excess of two electrons. Single qubit operations
are performed by modulating the gate-voltage difference between the
two dots, as well as through the coupling to an inhomogeneous
Overhauser field. For instance, calling the singlet states ``up'' and
triplet state ``down'' pseudospin states, we can write the following
pseudospin Hamiltonian
\begin{equation}
H_S = \left[ \frac{\partial J}{\partial \varepsilon} (\varepsilon)
  \right]\, e\eta\, \left( V_1 - V_2 \right), \sigma_z + H_{HF},
\label{eq:Hspin}
\end{equation}
where $H_{HF}$ describes the coupling to the Overhauser field and
$J(\varepsilon)$ is the effective exchange coupling. The latter can be
calculated in second-order perturbation theory,
\begin{equation}
J(\varepsilon) = 4t^2 \left[ \frac{1}{(U-\varepsilon)} +
  \frac{1}{(U+\varepsilon)} \right],
\end{equation}
with $t=\Delta/2$ denoting the interdot tunneling matrix element and
$U=E_{C1} = E_{C2}$ representing the dot charging energy (for the sake
of simplicity, we assume equal charging energy for both dots). The
values $\varepsilon = \pm U$ mark the transitions from ($N_1=1,N_2=1$)
to ($N_1=2,N_2=0$) or ($N_1=0,N_2=2$) states and the breakdown of the
perturbative expansion.

The first term in Eq. (\ref{eq:Hspin}) is quite similar to first term
in Eq. (\ref{eq:Hs}). Gate voltage fluctuations will couple to this
spin qubit similarly to the case of the DQD charge
qubit.\cite{marquardt05} Therefore, for small biases ($|\varepsilon|
\ll U$), we can study decoherence induced by electromagnetic
fluctuations in the spin qubit employing the same analysis developed
in the previous sections for the charge qubit. We note that several
studies of decoherence due to other mechanisms also present in these
qubits have been done.\cite{taylor,ribeiro,culcer09}

The decoherence rates will depend strongly on the qubit operation
point, given that the prefactor $|\partial J/\partial \varepsilon|$
appearing in Eq. (\ref{eq:Hspin}) varies rapidly with
$\varepsilon$. Qubit operations around this point require pulsing the
exchange coupling $J(\varepsilon)$ for a time interval $\tau_E$,
during which the qubit may be vulnerable to dephasing due to
electromagnetic fluctuations. An estimate of the corresponding
decoherence time can be obtained by using the curve $J(\varepsilon)$
plotted in Fig. 3d of Ref. \onlinecite{petta05}. Near $\varepsilon =
-1$ mV, one finds $|\partial J/\partial \varepsilon| \approx 10^{-3}$.
Since decoherence rates are proportional to the square of the
bath-qubit coupling, namely, $|\partial J/\partial \varepsilon|^2$ in
this case, we conclude that the decoherence times due to
electromagnetic fluctuations are $10^6$ larger than those found for
the DQD charge qubits, hence ranging from tens to hundreds of
millisecond. In practice, these times are much larger than $\tau_E$,
which is typically a few hundreds of picosecond. We can conclude that
gate-voltage fluctuations are also not a significant source of
decoherence for spin qubits.

\section{Conclusions}
\label{sec:conclusions}

In this paper, we have modeled noise introduced by gate-voltage
fluctuations in double-quantum-dot systems. We attempted to model the
circuits leading to the DQD in a way that put us as close to real
experimental values as possible, while still being able to estimate
all the relevant parameters and calculate decoherence rates and
quality factors.

We chose to place our noise sources in our gates because we believe
they give the largest contribution to decoherence during qubit
operations. For additional considerations, noise sources could also be
placed, for example, in the drain and source electrodes.

We have estimated the effect of fluctuations in the electrodes feeding
the quantum dots and shown the influence that parameters such as
temperature and inter-capacitive coupling between electrodes have on
decoherence in qubit operation. We have also shown that, similarly to
decoherence by phonon coupling, temperature degrades coherence in the
state superpositions, reinforcing the need for efficient refrigeration
of the leads. This effect can be explained analogously to the
radiation of a black body, which increases with temperature.

Contrary to what was initially expected, it was found that a stronger
inter-capacitive coupling between electrodes actually introduces
stronger decoherence in the qubit system. Thus, in order to mitigate
this effect, it is important to keep the leads gating each quantum dot
in the system as isolated as possible from each other.

There are a few possible refinements to the model presented here. One
such improvement includes adding the electrical resistance in the
leads, which in practice requires the use of a lossy transmission line
model for the effective circuit. It may also be important to take into
account the drag effect on the leads due to the proximity to the 2DEG.
This effect will change the effective circuit parameters, thus
influencing the calculation of relaxation and dephasing times.

We have found that electromagnetic fluctuations in DQD systems do not
introduce a dominating decoherence effect. The quality factors
calculated for our system at room temperature ($\sim 210$) are still
well above the $Q$ factors found in systems under the effect of phonon
coupling ($\sim 50$).\cite{sergueietaldoubledot05,wu04,Hu05,hentschel}
If we compare these results with the experimental results ($\sim 3-9$)
for $Q$ factors, the discrepancy is even
larger.\cite{hayashi03,petta04,gorman05}

The disagreement between theoretical estimates and measured
decoherence times in charge based DQD system leads us to believe that
there must be another possible noise source that accounts for the
short decoherence times observed in these systems. For instance, it
has been recently argued that electron-electron interactions can
enhance the effect of fluctuating background charges on the charge
qubits.\cite{abel08,lerner}

In order to identify the leading decoherence mechanism in charge-based
qubits, it would be very helpful if the dependence of the $Q$ factor
on the qubit operating frequency $\nu$ were measured. For instance,
for bosonic environments, this would yield the spectral function. With
this information in hand, one could perhaps trace back the physical
process underlying the decoherence mechanism. A candidate for such
source is the presence of fluctuating background charges trapped in
the insulating substrate or at the GaAS/GaAlAs interface.

\section{Acknowledgments}

We are grateful to A. Chang, R. Hanson, and J. Kycia for providing us
with information related to their experimental setups, as well as
W. Coish, A. Fowler, E. Novais, and J. M. Taylor for useful
discussions. This work was supported in part by the NSF under Grant
No. CCF 0523603 and by the Office of Naval Research. D.C.B.V. and
E.R.M. acknowledge partial support from the Interdisciplinary
Information Science and Technology Laboratory (I$^2$Lab) at
UCF. F.K.W. acknowledges support by NSERC through the Discovery Grants
program and through QuantumWorks.



\end{document}